\documentclass[11pt]{article}
\usepackage{amsfonts}
\usepackage{amsthm}
 \usepackage{bezier}

\usepackage{epic,eepic,amsmath,amssymb,color,graphicx}

\def\be{\begin{equation}}
\def\ee{\end{equation}}
\def\bea{\begin{eqnarray}}
\def\eea{\end{eqnarray}}

\newcommand{\sect}[1]{\setcounter{equation}{0}\section{#1}}

\newcommand{\bq}{\mathbf{q}}
\newcommand{\bp}{\mathbf{p}}

\newcommand{\btq}{ \mathbf{q}}
\newcommand{\btp}{ \mathbf{p}}

\newcommand{\tq}{ {q}}
\newcommand{\tp}{ {p}}

\newcommand{\bttq}{ \mathbf{\tilde q}}
\newcommand{\bttp}{ \mathbf{\tilde p}}

\newcommand{\ttq}{ {\tilde q}}
\newcommand{\ttp}{ {\tilde p}}

 \newcommand{\kk}{\kappa}
   \newcommand{\del}{\delta}
  \newcommand{\Om}{\Omega}
 \newcommand{\la}{\lambda}
\newcommand{\te}{\phi}

 \newcommand{\non}{ {\cal H}^{1:2}}

  \newcommand{\nonn}{ {\cal H}}

\newcommand{\ii}{ {\cal I}^{1:2}}

\newcommand{\iii}{ {\cal I}}

\newcommand{\ele}{ {\cal L}}

\newcommand{\Sk}{{\rm\ \!S}}            
\newcommand{\Ck}{{\rm\ \!C}}           
\newcommand{\Tk}{{\rm\ \!T}}

\def\1{\'{\i}}                           

  \def\>#1{{\mathbf#1}}

 \def\pp{P}

 \def\m{\mu}
\def\k{{\kappa}}

\begin{document}

\thispagestyle{empty}

\ 

 \vskip2cm

\begin{center}

 {\Large{\bf {A new integrable anisotropic oscillator on the
\\[6pt] two-dimensional 
sphere and the  hyperbolic plane}}}

\medskip 
\medskip 
\medskip

{\sc \'Angel Ballesteros$^1$, Alfonso Blasco$^1$, Francisco J.~Herranz$^1$ and Fabio Musso$^{2}$}

\medskip

{$^1$ Departamento de F\'\i sica,  Universidad de Burgos,
E-09001 Burgos, Spain  \\
$^2$ Dipartimento di Matematica e Fisica,  Universit\'a Roma Tre,
I-00146 Rome, Italy \\
}
\medskip 

\noindent
 E-mail: {\tt   angelb@ubu.es,  ablasco@ubu.es,  fjherranz@ubu.es, musso@fis.uniroma3.it}

\end{center}

  \medskip 
\bigskip
\bigskip

\begin{abstract} 
\noindent A new integrable generalization to the 2D sphere ${\mathbf S}^2$ and to the hyperbolic space ${\mathbf H}^2$ of the 2D Euclidean anisotropic oscillator Hamiltonian with Rosochatius (centrifugal) terms is presented, and its curved integral of the motion is shown to be quadratic in the momenta. In order to construct such a new integrable Hamiltonian $\nonn_\kk$, we will make use of a group theoretical approach in which the curvature $\k$ of the underlying space will be treated as an additional (contraction) parameter, and we will make extensive use of projective coordinates and their associated phase spaces. It turns out that when the oscillator parameters $\Om_1$ and $\Om_2$ are such that $\Om_2=4\Om_1$, the system turns out to be the well-known superintegrable $1:2$ oscillator on ${\mathbf S}^2$ and ${\mathbf H}^2$. Nevertheless, numerical integration of the trajectories of $\nonn_\kk$  suggests that for other values of the parameters $\Om_1$ and $\Om_2$ the system is not superintegrable. In this way, we support the conjecture that for each commensurate (and thus superintegrable) $m:n$  Euclidean oscillator there exists a two-parametric family of curved integrable (but not superintegrable) oscillators that turns out to be superintegrable only when the parameters are tuned to the $m:n$ commensurability condition. 

 \end{abstract}

\bigskip\bigskip\bigskip\bigskip


\bigskip


\newpage


\sect{Introduction}

In this paper we present a new integrable generalization of the 2D Euclidean anisotropic oscillator Hamiltonian with Rosochatius (centrifugal) terms
\be
{\cal H}=\frac 12 (p_1^2+p_2^2)+\Om_1 q_1^2+\Om_2 q_2^2 +\frac{\la_1}{q_1^2}+\frac{\la_2}{q_2^2},
\label{aa}
\ee
to the 2D sphere ${\mathbf S}^2$ and to the hyperbolic space ${\mathbf H}^2$. Such generalization $\nonn_\kk$  is shown to be integrable for any value of the set of parameters $\{\Om_1>0,\Om_2>0,\la_1,\la_2\}$, and the curved integral of the motion that provides the integrability of $\nonn_\kk$  turns out to be quadratic in the momenta.  

Evidently, for {\em any} value of the $\Om_1$ and $\Om_2$ parameters the Hamiltonian~\eqref{aa} is indeed separable and, as a consequence, integrable in the Liouville sense (note that the separability of the kinetic energy term will be lost when a non-zero curvature is introduced). On the other hand, the Euclidean system~\eqref{aa} is known to be superintegrable ({\em i.e.} possesing a third independent and globaly defined constant of the motion) only when the two frequencies $\omega_1$ and $\omega_2$, defined through $\omega_1^2=2\Omega_1 $ and $\omega_2^2=2\Omega_2 $,  are commensurate. Moreover, the two integrals of the motion for a given commensurate Euclidean oscillator are quadratic in the momenta only in the $1:1$ and $1:2$ cases, being of higher-order for any other superintegrable case (see~\cite{Perelomov, Jauch, Tempesta}). 

If order to generalize these results to a constant curvature scenario~\cite{Voz}, we will make use of a simultaneous construction of ${\mathbf S}^2$ and ${\mathbf H}^2$ as homogeneous spaces of the one-parametric ${\rm SO}_\kk(3)$  Lie group, in which the curvature $\kk$ will be introduced as a contraction parameter~\cite{Yaglom, Groma, CK2}. Within this approach, an integrable system on ${\mathbf S}^2$ and ${\mathbf H}^2$ will be called an {\em anisotropic curved oscillator} with centrifugal terms if its Euclidean limit $\k\to 0$ gives~\eqref{aa}. 

We have to recall that the classification of all possible superintegrable systems on ${\mathbf S}^2$ and ${\mathbf H}^2$ with quadratic integrals of the motion was presented in \cite{RS}. Among them we find only two curved superintegrable oscillator potentials: the Higgs oscillator, whose Euclidean limit is the $1:1$ isotropic oscillator ($\Om_2=\Om_1$, with $\la_1$ and $\la_2$ arbitrary), together with an anisotropic oscillator whose zero curvature limit is the superintegrable Euclidean $1:2$ oscillator ($\Om_2=4\,\Om_1$, with $\la_2=0$ and $\la_1$ arbitrary). 

To the best of our knowledge, no other commensurate Euclidean oscillator has been so far generalized to the curved case.
Moreover, only recently a first integrable (but not superintegrable) generalization of~\eqref{aa} for arbitrary values of $\{\Om_1,\Om_2,\la_1,\la_2\}$ has been presented in~\cite{Non}. This curved Hamiltonian gives rise to the superintegrable Higgs oscillator when $\Om_1=\Om_2$, and therefore it can be interpreted as the `anisotropic generalization' of the Higgs oscillator. Due to this fact, and taking also into account that the curved superintegrable $1:2$ oscillator was shown in~\cite{Non} to be another different system,  we then conjectured that for each commensurate $m:n$ Euclidean oscillator there should exist a {\em different} two-parametric $(\Om_1,\Om_2)$ curved anisotropic oscillator system that should provide the corresponding curved superintegrable $m:n$ system when appropriate $(\Om_1,\Om_2)$ values are considered. 

The aim of this paper is to support this conjecture by presenting a new curved anisotropic oscillator $\nonn_\kk$    that
\begin{itemize}
\item is Liouville integrable (but not superintegrable) for arbitrary values of $\{\Om_1,\Om_2,\la_1,\la_2\}$,
\item it reduces to the curved superintegrable $1:2$ oscillator when $\Om_2=4\,\Om_1$,
\item and it gives~\eqref{aa} under the Euclidean limit $\kk\to 0$. 
\end{itemize}
The fact that $\nonn_\kk$ does not seem  to be superintegrable whenever $\Om_2\neq4\,\Om_1$ will be studied through numerical integration of its bounded trajectories, that turn out to be non-closed ones for all the cases so far studied. To this respect, the description of the system in terms of projective canonical variables will be helpful, and the plot on the Poincar\'e disk of the potential defining $\nonn_\kk$ provides an unexpected geometric footprint of superintegrability.

The paper is structured as follows. In the next Section we present the Cayley--Klein approach to ${\mathbf S}^2$,  ${\mathbf H}^2$ and ${\mathbf E}^2$ that will provide a simultaneous description of the  dynamics on these three spaces in terms of the curvature parameter $\kk$. In particular, the expression of the kinetic energy will be presented in ambient and geodesic polar coordinates. In Section 3 the    Poincar\'e and Beltrami projective coordinates and their associated phase spaces will be introduced and related with the ambient and geodesic polar descriptions. In this way, Sections 2 and 3 provide a self-contained approach to the geometric dynamics on ${\mathbf S}^2$ and ${\mathbf H}^2$ that completes the description already presented in~\cite{RS,Non}.  In Section 4 the known superintegrable $1:1$ and $1:2$ oscillator systems on ${\mathbf S}^2$ and  ${\mathbf H}^2$ are revisited, and the usefulness of the projective coordinates in this type of curved integrability problems will be illustrated. Section 5 includes the main new result of the paper: the explicit form of the integrable anisotropic oscillator $\nonn_\kk$ that leads to the superintegrable curved oscillator $1:2$ in the specific $\Om_2=4\,\Om_1$ case. The non-superintegrability of $\nonn_\kk$ for generic values of $\Om_1$ and $\Om_2$ is illustrated in Section 6 through the numerical integration of a selected sample of trajectories of the system, and the use of projective variables turns out to be also helpful  from this numerical perspective. Finally, some comments and open problems close the paper.


\sect{Geometry and geodesic dynamics on ${\mathbf S}^2$ and ${\mathbf H}^2$}

 To start with, we provide a complete group theoretical approach of ${\mathbf S}^2$ and ${\mathbf H}^2$ which leads  to a very natural description of the 3D ambient variables for these spaces. In particular, we present all the relevant dynamical quantities in terms of the latter variables, thus including the ambient momenta,  which was lacking in previous works on the subject (see~\cite{RS,Non} and references therein) and turns out to be useful. In particular, we explicitly provide the case of geodesic polar coordinates.


\subsection{Vector model and ambient canonical  variables}

Let us consider a one-parametric family of 3D real Lie algebras (which is contained within the family of the so-called 2D orthogonal Cayley--Klein
   algebras~\cite{Yaglom, Groma, CK2}) that we denote collectively as $\mathfrak{so}_\kk(3)$ where
  $\k$ is  a real contraction parameter.   The   Lie brackets of
 $\mathfrak{so}_\kk(3)$ in the basis spanned by
$\langle J_{01}, J_{02}, J_{12}\rangle$ are given by 
\be
  [J_{12},J_{01}]=J_{02},\qquad [J_{12},J_{02}]=-J_{01},\qquad [J_{01},J_{02}]=\kk J_{12}  . \label{ca}
 \ee
 The Casimir invariant, coming from the  Killing--Cartan form, reads
\be
  {\cal C}=J_{01}^2+J_{02}^2+\kk J_{12}^2.
 \label{caa}
 \ee
 Notice that by redefining appropriately the Lie generators the real parameter $\k$ can be rescaled to either $+1$, 0 or $-1$.
In particular, putting $\k=0$ is   equivalent to applying an In\"on\"u--Wigner contraction~\cite{IW}. In fact, 
the involutive
automorphism  defined by
$$
\Theta  :  \ J_{12}\to J_{12},\quad J_{0i}\to -J_{0i},\quad i=1,2,
$$
generates a $\mathbb Z_2$-grading of   $\mathfrak{so}_\kk(3)$  in such a manner that $\k$ is  a
graded contraction parameter~\cite{Montigny}. This automorphism  gives  rise to the following Cartan
decomposition:
$$
 \mathfrak{so}_\kk(3) ={\mathfrak{h}}  \oplus  {\mathfrak{p}} ,\qquad 
{\mathfrak{h}  }=\langle J_{12} \rangle=\mathfrak{so}(2) ,\qquad
{\mathfrak{p}  }=\langle J_{01},J_{02} \rangle  .
$$

Now, the   family   of the  three classical  2D Riemannian  symmetric homogeneous
spaces  with constant Gaussian curvature   $\k$  is defined by  the quotient ${\mathbf
S}^2_\k \equiv {\rm  SO}_{\k}(3)/{\rm  SO}(2) $ where  the Lie groups  $H = {\rm  SO}(2) $  and ${\rm SO}_{\k}(3)$ have   ${\mathfrak{h}} $ and  $\mathfrak{so}_\kk(3)$,   respectively, as Lie algebras. Namely,
 $$
\begin{array}{lll}
\kk>0:\ \mbox{2D Sphere}&\quad \kk=0:\ \mbox{Euclidean plane}&\quad \kk<0:\ \mbox{2D Hyperbolic space}\\[2pt]
{\mathbf
S}^2_+ \equiv  {\mathbf S}^2={\rm SO}(3)/{\rm  SO}(2)&\quad  {\mathbf
S}^2_0 \equiv {\mathbf E}^2={\rm  ISO}(2)/{\rm  SO}(2)&\quad {\mathbf
S}^2_- \equiv  {\mathbf H}^2= {\rm  SO}(2,1)/{\rm SO}(2)
\end{array}
$$

The {\em vector representation} of  $\mathfrak{so}_\kk(3)$ is  given by the following $3\times 3$ real
matrices:
\be
J_{01}=\left(\begin{array}{ccc}
.& -\k & .\cr
1& . & . \cr
.& . & . 
\end{array}\right),
\qquad 
J_{02}=\left(\begin{array}{ccc}
.& . & -\k  \cr
.& . &  .\cr
1& . & .
\end{array}\right) ,\qquad 
 J_{12}=\left(\begin{array}{ccc}
.& . & . \cr
.& . & -1 \cr
.& 1 & .
\end{array}\right).
\label{ccaa}
\ee
 Their exponentials lead to the corresponding one-parametric subgroups of
${\rm  {SO}}_{\k }(3)$:
\begin{equation}
\begin{array}{l}
{\rm e}^{\alpha J_{01}}=\left(\begin{array}{cccc}
\Ck_{\k}(\alpha)& -\k \Sk_{\k}(\alpha)& . \cr
\Sk_{\k}(\alpha)& \Ck_{\k}(\alpha) & . \cr
.& . & 1 &  
\end{array}\right) ,\quad  {\rm e}^{\gamma J_{12}}=\left(\begin{array}{cccc}
1& . & . \cr
.& \cos  \gamma& -\sin \gamma\cr
.& \sin  \gamma &\cos \gamma \cr
\end{array}\right) ,\\[25pt]
{\rm e}^{\beta J_{02}}=\left(\begin{array}{ccc}
 \Ck_{\k}(\beta)& . &\ -\k   \Sk_{\k}(\beta) \cr
\!\!\!.& 1 & . \cr
\!\!\!\Sk_{\k}(\beta)& . & \Ck_{\k}(\beta) 
 \end{array}\right) ,\quad
 \end{array}
\label{cb}
\end{equation}
where we have introduced the $\k$-dependent cosine and sine functions defined
by~\cite{CK2}
\begin{equation}
\Ck_{\k}(x) =\sum_{l=0}^{\infty}(-\k)^l\frac{x^{2l}} 
{(2l)!}=\left\{
\begin{array}{ll}
  \cos {\sqrt{\k}\, x} &\quad  \k>0 \\ 
\qquad 1  &\quad
  \k=0 \\
\cosh {\sqrt{-\k}\, x} &\quad   \k<0 
\end{array}\right.  ,
\nonumber
\end{equation}
\begin{equation}
   \Sk{_\k}(x) =\sum_{l=0}^{\infty}(-\k)^l\frac{x^{2l+1}}{ (2l+1)!}
= \left\{
\begin{array}{ll}
  \frac{1}{\sqrt{\k}} \sin {\sqrt{\k}\, x} &\quad  \k>0 \\ 
\qquad x  &\quad
  \k=0 \\ 
\frac{1}{\sqrt{-\k}} \sinh {\sqrt{-\k}\, x} &\quad  \k<0 
\end{array}\right.  .
\nonumber
\end{equation}
 The $\k$-tangent is defined
as $\Tk_\k(x)=\Sk_\k(x)/\Ck_\k(x)$, and 
different useful relations  involving these $\k$-functions   can be found in~\cite{RS, trigo, conf}, for instance 
\[
\Ck^2_\k(x)+\k\,\Sk^2_\k(x)=1,\quad \frac{ {\rm d}}
{{\rm d} x}\Ck_\k(x)=-\k\,\Sk_\k(x),\quad 
\frac{ {\rm d}}
{{\rm d} x}\Sk_\k(x)= \Ck_\k(x)  ,\quad 
\frac{ {\rm d}}
{{\rm d} x}\Tk_\k(x)=  \frac{1}{\Ck^2_\k(x) }.
\]

Under the   matrix    representations (\ref{ccaa}) and (\ref{cb}), the following relations hold 
\[
X^T \mathbf I_{\k}+\mathbf I_{\k} X=0,\quad  X\in {\mathfrak {so}}_{\k}(3),\qquad
Y^T \mathbf I_{\k} Y=\mathbf I_{\k} ,\quad  Y\in {\rm SO}_{\k}(3),
\]
($X^T$ denotes the transpose of $X$) with respect to the bilinear form
\be
\mathbf I_{\k}={\rm diag}(+1,\k,\k ) .
\label{ccdd}
\ee
 Therefore
${\rm SO}_{\k}(3)$ is a group of isometries of  $\mathbf I_{\k}$ acting on
a 3D {\em ambient}  linear  space 
$\mathbb R^3=(x_0,x_1,x_2)$ through matrix multiplication. The origin $O$ in ${\mathbf
S}^2_\k $ has ambient coordinates $O =(1,0,0)$ and this point is invariant under
the subgroup $  {\rm SO} (2)=\langle J_{12}\rangle$ (see (\ref{cb})). The orbit
of $O$ corresponds to the homogeneous space ${\mathbf
S}^2_\k $, which is contained
in the `sphere'
\begin{equation}
\Sigma_\k\equiv x_0^2+\k \left(x_1^2+   x_2^2\right)  =1 ,
\label{cde}
\end{equation}
determined by $\mathbf I_{\k}$ (\ref{ccdd}).  Hence when $\k>0$ we recover a proper sphere but when $\k<0$ we find the two-sheeted hyperboloid. In fact, if we write $\k=\pm 1/R^2$ where $R$ is the radius of the space, then the contraction $\k\to 0$ corresponds to the flat limit $R\to \infty$, that gives rise to two Euclidean planes $x_0=\pm 1$ with Cartesian coordinates $(x_1,x_2)$. Hereafter when dealing with the   hyperbolic and Euclidean spaces we shall consider  the upper sheet of the hyperboloid with $x_0\ge 1$ and the Euclidean plane   with $x_0=+1$.

The {\em ambient coordinates} $(x_0,x_1,x_2)$, subjected to the constraint
(\ref{cde}),  are also called {\it Weierstrass coordinates}.  In these variables the metric on ${\mathbf
S}^2_\k$
follows from    the flat ambient metric in $\mathbb R^{3}$ divided by the curvature $\k$ and
restricted to $\Sigma_\k$:
\begin{equation}
{\rm d} s^2=\left.\frac {1}{\k}
\left({\rm d} x_0^2+   \k \left( {\rm d} x_1^2+   {\rm d} x_2^2 \right)
\right)\right|_{\Sigma_\k}  =    \frac{\k\left(x_1{\rm d} x_1 + x_2{\rm d} x_2 \right)^2}{1-  \k \left(x_1^2+   x_2^2\right)}+  {\rm d} x_1^2+   {\rm d} x_2^2 .
\label{ce}
\end{equation}

A {\em differential realization} of the Lie algebra $\mathfrak{so}_{\k }(3)$ (\ref{ca}) in terms of first-order vector
fields in the ambient coordinates, can be easily deduced from the vector representation (\ref{ccaa}) and reads 
$$
J_{01}=\k \, x_1\frac{\partial}{\partial{x_0}} -x_0\frac{\partial}{\partial{x_1}}   ,\qquad
J_{02}=\k  \,x_2 \frac{\partial}{\partial{x_0}}   -x_0\frac{\partial}{\partial{x_2}} ,\qquad 
J_{12}=  x_2\frac{\partial}{\partial{x_1}}   -x_1\frac{\partial}{\partial{x_2}}   .
$$
 From it,  a {\em phase space (symplectic) realization} of the Lie generators of
$\mathfrak{so}_{\k }(3)$ in terms of Weierstrass coordinates
$x_\mu$ and their conjugate momenta $\pp_\mu$  $(\mu=0,1,2)$ is obtained by setting
$\partial_\mu\to -\pp_\mu$:
\begin{equation}
J_{01}=x_0 \pp_1-\k\, x_1 \pp_0 ,\qquad
J_{02}= x_0 \pp_2-\k\, x_2 \pp_0,\qquad 
J_{12}=  x_1 \pp_2-x_2 \pp_1 .
\label{eda}
\end{equation}
As we will see in the sequel, these three functions will provide the essential building blocks for the integrals of motion of all the Hamiltonians we will deal with in this paper.

From  the metric (\ref{ce}), the free Lagrangian ${\cal L}_\k$ defining the geodesic motion of a particle with unit mass in  the 2D space ${\mathbf
S}^2_\k $ with ambient velocities
$\dot x_\mu$ is obtained:
\be
{\cal L}_\k=\frac 1{2\k} \left. \left( \dot x_0^2+   \k \left(  \dot x_1^2+  \dot  x_2^2 \right)\right)
 \right |_{\Sigma_\k}  =    \frac{\k\left(x_1\dot  x_1 + x_2\dot  x_2 \right)^2}{2\left(1-  \k \left(x_1^2+   x_2^2\right)\right)}+  \frac 12 \left( \dot  x_1^2+   \dot  x_2^2\right),
\label{css}
\ee
and the corresponding   momenta $\pp_\mu=\partial {\cal L}_\k/\partial \dot x_\mu$  $(\mu=0,1,2)$  turn out to be
 \begin{equation}
\pp_0=\dot x_0/\k,\qquad \pp_1=\dot x_1,\qquad \pp_2= \dot x_2,
\label{edb}
\end{equation}
in a way consistent with  (\ref{cde}), which means that
$$
\Sigma_\k\equiv x_0\pp_0+x_1\pp_1+x_2\pp_2=0 .
$$
Hence the kinetic energy ${\cal T}_\k$ in ambient coordinates reads
\be
{\cal T}_\k=\frac 1{2 } \left. \left(\k\,   \pp_0^2+       \pp_1^2+     \pp_2^2  \right)
 \right |_{\Sigma_\k}  =    \frac{\k\left(x_1   \pp_1 + x_2   \pp_2 \right)^2}{2\left(1-  \k \left(x_1^2+   x_2^2\right)\right)}+  \frac 12 \left(  \pp_1^2+      \pp_2^2\right).
 \label{Tkambient}
\ee

Evidently, the ambient coordinates $x_\mu$ can be
parametrized in terms of  {\em two intrinsic} quantities in different
ways.  For our purposes, we shall introduce the so-called  geodesic polar coordinates (as a generalization of the usual Euclidean polar ones) and two sets of projective variables.


\subsection{Geodesic polar variables}

Let us now consider  a point $Q\in {\mathbf
S}^2_\k $  with ambient coordinates $(x_0,x_1,x_2)$. The {\em geodesic polar coordinates} $(r ,\phi)$   are
defined~\cite{conf} through the following action of two of the one-parametric subgroups (\ref{cb}) onto the
origin
$O=(1,0,0)$, namely
\begin{equation}
\begin{array}{l}
Q(x_0,x_1,x_2)\equiv Q(r,\phi)=\exp\{\phi J_{12}\} \exp\{r
J_{01}\}O,
\end{array}
\nonumber
\end{equation}
which means that
\begin{equation}
\left(\begin{array}{c}
x_0\cr
x_1\cr
x_2 
\end{array}\right)=
\left(\begin{array}{c}
\Ck_{\k}(r)\cr
\Sk_{\k}(r)\cos  \phi\cr
\Sk_{\k}(r)\sin  \phi
\end{array}\right).
\label{da}
\end{equation}

Let  now  $l_1$ and $l_2$ be  two base geodesics in ${\mathbf
S}^2_\k$ which are orthogonal at the origin $O$. 
Then the  `radial' coordinate $r$ is the geodesic  distance between $Q$ and   $O$
measured along  the geodesic $l$ that joins both points,  while $\phi$ is the angle which determines the orientation of $l$ with respect to the base geodesic $l_1$  (see~\cite{Non} for details). 
Hence  each   `translation' generator $J_{0i}$ moves   $O$  along the base geodesic  $l_i$ $(i=1,2)$ while the `rotation' one $J_{12}$ leaves $O$ invariant.
According to each specific Riemannian  space  ${\mathbf
S}^2_\k$ we find that:

 \begin{itemize}
\itemsep=0pt
\item  In the sphere ${\mathbf
S}^2_+ \equiv {\mathbf
S}^2  $ we have $\k=1/R^2>0$ and the `radial' coordinate $r$ has dimensions of {\it length}, $[r]=[R]$. Notice   that   the 
dimensionless  coordinate $r/R$, which is   an ordinary angle, is usually
taken  instead of $r$ (see,
e.g.,~\cite{Pogosyan}). In this case $r\in[0, \pi/\sqrt{\k})$ while $\phi \in [0,2\pi)$.

 \item  In the hyperbolic or Lobachevski space ${\mathbf
S}^2_- \equiv {\mathbf
H}^2  $ with $\k=-1/R^2<0$,   the `radial' coordinate $r$ has also dimensions of {\it length} but now  $r\in[0, +\infty)$ and $\phi \in [0,2\pi)$.

 \item Finally, in the  flat (contracted) Euclidean plane ${\mathbf
S}^2_0 \equiv {\mathbf
E}^2  $ with $\k=0$ ($R\to \infty)$,  we recover the usual polar coordinates such that $r\in[0, +\infty)$ and $\phi \in [0,2\pi)$.

\end{itemize}

It is straightforward to check that, by introducing  (\ref{da}) in the ambient metric (\ref{ce}) and in the free Lagrangian (\ref{css}) we obtain
$$
{\rm d} s^2= 
 {\rm d} r^2+    \Sk_{\k}^2(r)  {\rm d}\phi^2,\qquad {\cal L}_\k=\frac 12\left(\dot r^2+   \Sk_{\k}^2(r) \dot \phi^2 \right)  ,
$$
and the conjugate momenta to the coordinates $(r,\te)$ turn out to be 
\be
p_r= \dot r,\qquad p_\phi =\Sk_{\k}^2(r) \dot \phi .
\label{dcc}
\ee
Hence the kinetic energy reads
\be
 {\cal T}_\k=\frac 12\left(p_r^2+  \frac{p_\phi^2}{ \Sk_{\k}^2(r)}   \right)  .
\label{dc}
\ee

Furthermore, by  substituting   (\ref{dcc})  within  (\ref{edb})  we obtain   the relationships between the
ambient and the geodesic polar momenta, $\pp_\mu(r,\phi)$, which are summarized in table 1 
together with a symplectic realization of the $\mathfrak{so}_\kk(3)$ generators  (\ref{eda})  in terms of geodesic polar
variables, $J_{\mu\nu}(r,\phi,p_r,p_\phi)$.
It is worth remarking that the kinetic energy $ {\cal T}_\k\equiv\frac 12 {\cal C}$ (\ref{dc})  can also be recovered by computing the symplectic realization of the Casimir function $ {\cal C}$ (\ref{caa}) of $\mathfrak{so}_\kk(3)$.

 We recall that these geodesic polar coordinates were the ones used (together with the  the so-called geodesic parallel ones)  in the classification of the 2D {\em superintegrable} systems with {\em quadratic} integrals of motion in the momenta on the 2D sphere and      hyperbolic space performed in~\cite{RS}, in which both the curved isotropic oscillator (Higgs oscillator) and the $1:2$ superintegrable oscillator arised. 
 Moreover, these coordinates  were also used  in the construction  of the curved anisotropic $1:1$ oscillator presented in~\cite{Non}.


\begin{table}[t]

 {\footnotesize{

\caption{{Expressions for the ambient variables $(x_\mu,\pp_\mu)$, the free Hamiltonian  $ {\cal T}_\k$ and  the symplectic realization of the Lie--Poisson generators $J_{\mu\nu}$  of $\mathfrak{so}_\kk(3)$ in terms of geodesic polar, Poincar\'e and Beltrami canonical variables. The specific expressions for ${\mathbf S}^2$,  ${\mathbf H}^2$ and ${\mathbf E}^2$ are obtained when $\kk>0, \kk<0$ and $\kk=0$, respectively.}}
\label{Table1}
 \begin{center}
\noindent
\begin{tabular}{llll}
\hline

\hline
\\[-0.2cm]
\multicolumn{1}{l}{ }&
\multicolumn{1}{l}{Polar variables $(r,\phi)$}&
\multicolumn{1}{l}{Poincar\'e variables $(\bttq, \bttp)$}
&\multicolumn{1}{l}{Beltrami variables $(\btq, \btp)$}\\[0.2cm]
\hline

\hline

\\[-0.2cm] 
$x_0$&$= \Ck_{\k}(r)$&  $=\displaystyle{  \frac{1- \kk \bttq^2}{ {1+ \kk \bttq^2}}    } $&$=\displaystyle{\frac{1}{ ({1+ \kk \btq^2})^{1/2}  }      }  $\\ [0.3cm]
$x_1$&$=\Sk_{\k}(r)\cos  \phi$&  $=\displaystyle{  \frac{2 \ttq_1}{ {1+ \kk \bttq^2}}     }  $&$=\displaystyle{ \frac{\tq_1}{ ({1+ \kk \btq^2})^{1/2}  }     }  $\\ [0.3cm]
$x_2$&$ =\Sk_{\k}(r)\sin  \phi$&  $=\displaystyle{   \frac{2 \ttq_2}{ {1+ \kk \bttq^2}}   }  $&$ =\displaystyle{\frac{\tq_2}{ ({1+ \kk \btq^2})^{1/2}  }      } $\\[0.5cm]
$\pp_0$&$ =- \Sk_{\k}(r)\, p_r$&  $ =\displaystyle{-(\bttq\cdot\bttp)  } $&$ =\displaystyle{ -\sqrt{1+\k\btq^2}\,(\btq\cdot \btp) } $\\[0.1cm]
$\pp_1$&$= \displaystyle{ \Ck_{\k}(r)\cos\phi\, p_r-\frac{\sin\phi}{\Sk_{\k}(r)}\,p_\phi}$&  $ =\displaystyle{ \tfrac 12 \left[(1+\k \bttq^2) \ttp_1 - 2 \k (\bttq\cdot \bttp) \ttq_1     \right ] } $&$=\displaystyle{ \sqrt{1+\k\btq^2}\, \tp_1 }  $\\ 
$\pp_2$&$= \displaystyle{ \Ck_{\k}(r)\sin\phi\, p_r+\frac{\cos\phi}{\Sk_{\k}(r)}\,p_\phi}$&  $=\displaystyle{\tfrac 12 \left[(1+\k \bttq^2) \ttp_2 - 2 \k (\bttq\cdot \bttp) \ttq_2     \right ]  }  $&$=\displaystyle{ \sqrt{1+\k\btq^2}\, \tp_2  }  $\\[0.3cm]
\hline
\\[-0.2cm]
 $ {\cal T}_\k$&$= \displaystyle{ \frac 12\left(p_r^2+  \frac{p_\phi^2}{ \Sk_{\k}^2(r)}   \right)  }   $&  $=\displaystyle{ \frac 18 \left(1+\kk \bttq^2\right)^2  \bttp^2 }  $&$ =\displaystyle{ \frac 12 \left(1+\kk \btq^2\right) \left(\btp^2+\kk(\btq\cdot\btp)^2 \right) } $\\[0.4cm]
\hline
\\[-0.2cm]
 $J_{01}$&$=  \displaystyle{  \cos\te\,p_r-\frac{\sin\te}{ \Tk_\kk(r) }\, p_\te}$&  $ =\displaystyle{  \tfrac 12 \left[ \ttp_1 +\k   (\bttq\cdot \bttp) \ttq_1 +\k\, \ttq_2 J_{12} \right ] } $&$ =\displaystyle{  \tp_1+\kk (\btq\cdot\btp) \tq_1 } $\\[0.3cm]
$ J_{02}$&$=  \displaystyle{ \sin\te\,p_r+\frac{\cos\te}{\Tk_\kk(r)}\, p_\te}$&  $=\displaystyle{   \tfrac 12 \left[ \ttp_2 +\k   (\bttq\cdot \bttp) \ttq_2 -\k\, \ttq_1 J_{12} \right ]   }  $&$ =\displaystyle{  \tp_2+\kk (\btq\cdot\btp) \tq_2 } $\\[0.3cm]
$ J_{12}$&$=  \displaystyle{  p_\te}$&  $=\displaystyle{  \ttq_1 \ttp_2 - \ttq_2 \ttp_1 } $&$=\displaystyle{ \tq_1 \tp_2 - \tq_2 \tp_1 }  $\\[0.2cm]
\hline

\hline

\end{tabular}
\end{center}
}}
 \end{table}



\sect{Projective coordinates and phase spaces}

 In this section we study in detail the two sets of canonical projective variables  that are well adapted to both the sphere and, specially, to the hyperbolic or  Lobachevsky plane: the Poincar\'e and Beltrami variables. Notice that the Poincar\'e ones were not considered in~\cite{Non} and, moreover, we here describe in detail the domain of both of them according to the value of the curvature $\k$. As we will see in the sequel, such analysis will be essential when computing, through numerical integration, the trajectories of the proposed Hamiltonians on these two curved spaces.


\subsection{Poincar\'e canonical variables}

Let us    consider the stereographic projection~\cite{Doubrovine}  with `south' pole  
$(-1,0,0)$  from the ambient coordinates   $ (x_0 ,x_1,x_2)\in \mathbb R^{3}$
to the {\em  Poincar\'e} ones  $ (\ttq_1,\ttq_2)\in \mathbb R^2$. This projection maps any point $(x_0 ,x_1,x_2)\in\Sigma_\k$  (\ref{cde})  through
$$
(x_0 ,x_1,x_2) \longrightarrow (-1,0,0)+\lambda\,
(1,\ttq_1,\ttq_2) .
$$
Hence we find that
\be
\lambda=\frac{2}{ {1+ \kk \bttq^2}},\qquad
 x_0= \lambda -1 =\frac{1- \kk \bttq^2}{ {1+ \kk \bttq^2}}   ,\qquad 
\>x=\lambda\, \bttq=\frac{2\bttq}{ {1+ \kk \bttq^2}} ,
\label{Poinc}
\ee
so that
$$
    \bttq=\frac{\>x}{1+x_0}, \qquad \bttq^2=\frac{1-x_0}{\k(1+ x_0)} ,
$$
where hereafter for any   object  with two components, say $\mathbf {a}=(a_1,a_2)$ and $\mathbf {b}=(b_1,b_2)$ we   denote 
$$
\mathbf {a}^2=a_1^2+a_2^2,  \qquad  |\mathbf {a}|=\sqrt{a_1^2+a_2^2}   , \qquad \mathbf {a}\cdot \mathbf {b}=a_1 b_1+ a_2 b_2 .
$$
Consequenlty, this projection is well defined for any point $Q \in \Sigma_\k$ except for the south pole $(-1,0,0)$ which goes to $\infty$ in both the sphere and the hyperbolic space. Notice that the ambient origin (`north' pole)  $O=(1,0,0)\in \Sigma_\k$  goes to the origin $\bttq=(0,0)$ of the 2D (projective) space ${\mathbf
S}^2_\k$.  In particular we find that:

 \begin{itemize}
\itemsep=0pt 
\item  In the sphere with $\k=1/R^2>0$,   $\bttq\in(-\infty,+\infty)$ and (\ref{Poinc}) maps  the equator  with $x_0=0$, {\em i.e.} $\>x^2=1/\k=R^2$, onto   the circle $\bttq^2=1/\k=R^2$, the northern hemisphere with $x_0>0$ onto the region inside that circle, $\bttq^2<1/\k$,  and  the southern one with $x_0<0$ onto   the outside region, $\bttq^2>1/\k$.

 \item  In the hyperbolic  space  with $\k=   -1/R^2<0$ and   such that $x_0\ge 1$ (the upper sheet of the hyperboloid) it is verified that
$\bttq\in \left[-1/\sqrt{|\k|}, + 1/\sqrt{|\k|} \right]$ and 
 $$
 \bttq^2= \frac{1}{|\k|}\left( \frac {x_0-1}{x_0+1}\right)\le R^2 ,
 $$
 which  is  just the  {\em Poincar\'e disk}. The points at the infinity in the hyperboloid correspond to the circle
 $ \bttq^2 =  {1}/{|\k|}=R^2  $
  (take $x_0\to +\infty$).

 \item In the  Euclidean plane with $\k=0$ ($R\to \infty)$, the Poincar\'e coordinates are {\em proportional} to the Cartesian ones $\>x \equiv\bq= 2\bttq$.

\end{itemize}

If we substitute (\ref{Poinc}) in the ambient metric (\ref{ce}) and   free Lagrangian (\ref{css}) we get the conjugate Poncar\'e momenta $\bttp$, namely
$$
{\rm d} s^2= \frac{4\,{\rm d}\bttq^2}{(1+\k\bttq^2)^2} ,\qquad {\cal L}_\k=\frac  {   2  \dot{   {\rm \tilde{ \bf q}  }    }^2}{  (1+\k\bttq^2)^2 }  ,\qquad \bttp= \frac  {4  \dot{   {\rm \tilde{ \bf q}  }    }   }{  (1+\k\bttq^2)^2 } .
$$
Proceeding similarly as in the previous Section,  the Poincar\'e symplectic realization of $\mathfrak{so}_\kk(3)$ can be computed. The final result is summarized in table 1. Notice that the factor 1/8 (instead of 1/2) in ${\cal T}_\k$  (also in ${\cal L}_\k$) is fully consistent with the fact that, under the $\k\to 0$ limit, Poincar\'e coordinates are twice the Euclidean ones.

It is also important to stress that Poincar\'e variables lead to a conformally flat diagonal metric with conformal factor $f$ such that
$$
    {\rm d} s^2=  f(|\bttq|)^2  {\rm d}\bttq^2,\qquad  f(|\bttq|)= \frac{2
}{1+\k\bttq^2} .
$$
Therefore, free motion will be described by the kinetic energy Hamiltonian
$$
{\cal T}_\k= \frac 1{2 f(|\bttq|)^2} \,\bttp^2 .
$$
Once more, on the Euclidean plane with $\k=0$ these espressions reduce to
$$
   f(|\bttq|)=2,\qquad  {\rm d} s^2=4 \, {\rm d} \bttq^2  =  {\rm d} \bq^2 =  {\rm d} {\>x}^2  ,\qquad   {\cal T}_0= \tfrac 1{8}\, \bttp^2  = \tfrac 12\,  \>p^2= \tfrac 12\,  \>P^2 ,
$$
since we have the relations $\bq\equiv \>x=2\bttq$ and $\bp\equiv\>\pp=\frac 12 \bttp$.


\subsection{Beltrami canonical variables}

The {\em Beltrami} projective coordinates $ (\tq_1,\tq_2)\in \mathbb R^2$  are  defined through
 the central projection with pole  
$(0,0,0)\in \mathbb R^{3}$ of a point $ (x_0, x_1,x_2)\in\Sigma_\k$. Explicitly, 
$$
(x_0, x_1,x_2)\longrightarrow (0,0,0)+\m\,
(1,\tq_1,\tq_2).
$$
This is tantamount to say
\be
\m=\frac{1}{\sqrt{1+ \kk \btq^2}},\qquad
 x_0=\m ,\qquad 
\>x=\m\, \btq=\frac{\btq}{\sqrt{1+ \k \btq^2}}.
\label{Beltr}
\ee
Hence
$$
\btq=\frac{\>x}{x_0},\qquad \btq^2= \frac{1-x_0^2}{\k\, x_0^2} .
$$
The  ambient origin  $O=(1,0,0)\in \Sigma_\k$  goes to the origin $\btq=(0,0)$ in the projective ${\mathbf
S}^2_\k$, as it should be. However, the 
domain of $\btq$ does depend on the   value of the curvature $\k$ as follows:

 \begin{itemize}
\itemsep=0pt 
\item  In ${\mathbf
S}^2 $ with $\k=1/R^2>0$,    $\btq\in(-\infty,+\infty)$. However, since the points in the equator with $x_0=0$ go to infinity, the projection (\ref{Beltr}) has to be separately defined for the two hemispheres with $x_0>0$ and $x_0<0$.

 \item  In ${\mathbf
H}^2$  with $\k=   -1/R^2<0$ and    $x_0\ge 1$,  
$\btq\in \left[-R, + R \right]$ and 
 $$
 \btq^2= \frac{x_0^2-1}{|\k|  x_0^2 } \le R^2 ,
 $$
 which  is    the  {\em Poincar\'e disk} in Beltrami coordinates. In a similar way as in  Poincar\'e coordinates, the points at the infinity in the hyperboloid go to the circle
 $ \btq^2 =  {1}/{|\k|}=R^2  $.

 \item In ${\mathbf
E}^2$ ($\k=0$), the Beltrami coordinates are just the Cartesian ones $\>x = \btq$.

\end{itemize}

In table 1 we also summarize the Beltrami symplectic realization of the Lie--Poisson algebra $\mathfrak{so}_\kk(3)$. Note that in this case the kinetic energy Hamiltonian is given by
$$
{\cal T}_\k=\frac 12 \left(1+\kk \btq^2\right) \left(\btp^2+\kk(\btq\cdot\btp)^2 \right),
$$
which was used in~\cite{Non} for the construction of the anisotropic Higgs oscillator. As it can be appreciated, the three sets of canonical variables presented in table 1 provide three very different expressions for the kinetic energy. In particular, the two projective coordinates provide {\em polynomial} quantities, that will simplify the search of algebraic invariants.  
 


\sect{Two quadratically superintegrable curved oscillators}

Now let us use the abovementioned variables in order to review the only two curved superintegrable oscillators with integrals that are quadratic in the momenta, as it was shown in \cite{RS}. The first one is the so-called  Higgs oscillator~\cite{RS,Non, Higgs,Leemon,int,CRMVulpi,kiev,Ranran, Nerssesian}, which is a curved analogue of the Euclidean isotropic $1:1$ oscillator. In terms of ambient, geodesic polar, Poincar\'e and Beltrami canonical variables the Higgs Hamiltonian   with two Rosochatius potentials  is written, respectively, as
  \bea
  &&{\cal H}_\k^{1:1}= {\cal T}_\k+  \del \,\frac{\>x^2}{(1-\k \>x^2)}  +\frac{\la_1}{x_1^2} +\frac{\la_2}{x_2^2} \nonumber\\
  && \quad\quad\  = {\cal T}_\k+ \del \Tk^2_\kk(r) +\frac{\la_1}{\Sk^2_\kk(r) \cos^2\te}+\frac{\la_2}{\Sk^2_\kk(r) \sin^2\te} \nonumber
  \\
   &&\quad\quad\  =  {\cal T}_\k+  \del\,\frac{4 \bttq^2}{(1-\k\bttq^2)^2}  + \frac 14 {(1+\k\bttq^2)^2}   \left(  \frac{\la_1}{  \ttq_1^2} +\frac{\la_2} { \ttq_2^2}  \right)
      \nonumber
  \\
   &&\quad\quad\  =  {\cal T}_\k+  \del \bq^2 +
\left(1+\kk \bq^2\right)\left(\frac{\la_1}{q_1^2}+\frac{\la_2}{q_2^2} \right),
    \label{Higgsall}
  \eea
where the corresponding expressions for the kinetic energy ${\cal T}_\k$ are given in~\eqref{Tkambient} and in table 1.  The Euclidean limit leading to the  $1:1 $ oscillator with two Rosochatius terms is obtained when $\k\to 0$. The integrals of the motion for this Hamiltonian are explicitly given in~\cite{Non}. A glimpse on~\eqref{Higgsall} makes evident that Beltrami coordinates are the most suitable for both symbolic and numerical computations concerning this system.

The second quadratically superintegrable oscillator found in \cite{RS} is a curved version of the anisotropic Euclidean $1:2$ oscillator. In this case the Hamiltonian reads
  \bea
 &&\!\!\!\!  \!\!\!\!    {\cal H}_\k^{1:2}= {\cal T}_\k+ \del\, \frac{x_1^2}{(1-\k x_1^2)}+4\del\, \frac{x_0^2x_2^2}{(x_0^2+\k x_2^2)(x_0^2-\k x_2^2)^2}+\frac{\la_1}{x_1^2}  \nonumber\\
  &&\!\!\!\!  \!\!\!\!   \quad\quad\  = {\cal T}_\k+\del \,\frac{\Sk^2_\kk(r)\cos^2\te}{\left(1-\kk \Sk^2_\kk(r)\cos^2\te\right)}   
  \nonumber\\[2pt]
   && \qquad \quad  + 4\del\, \frac{\Tk^2_\kk(r)\sin^2\te}{\left(1-\kk \Sk^2_\kk(r)\cos^2\te\right)\left(1-\kk \Tk^2_\kk(r)\sin^2\te\right)^2} +\frac{\la_1}{\Sk^2_\kk(r) \cos^2\te} \nonumber\\
      &&\!\!\!\!  \!\!\!\!  \quad\quad\  =   {\cal T}_\k+ \del\,\frac{4 \ttq_1^2}{  (1+\k\bttq^2)^2-4\k \ttq_1^2}  \nonumber \\
  &&\qquad \quad + 4\del\,\frac{4\ttq_2^2(1+\k\bttq^2)^2(1-\k\bttq^2)^2}{    \left( (1-\k\bttq^2)^2+4\k \ttq_2^2 \right ) \left( (1-\k\bttq^2)^2-4\k \ttq_2^2 \right )^2  }  +\la_1\,\frac{(1+\k\bttq^2)^2}{4 \ttq_1^2}  \nonumber
  \\
   &&\!\!\!\!  \!\!\!\!  \quad\quad\  = {\cal T}_\k+  \frac{\del \, \tq_1^2}{(1+\kk  \tq_2^2)}
+\frac{4\del\,   (1+\kk \btq^2) \tq_2^2}{ (1+\kk  \tq_2^2)(1-\kk  \tq_2^2)^2  }  +\la_1\,\frac{(1+\kk \btq^2)}{ \tq_1^2} .
    \label{MSall}
  \eea
As expected, the $\k\to 0$ limit leads to the Euclidean  $1:2$ oscillator with   a single Rosochatius term.
For further purposes and for the sake of completeness, let us summarize the full superintegrability properties of this curved  $1:2$ oscillator in terms of Beltrami variables.

\medskip

\noindent
{\bf Proposition 1.} \cite{Non}  {\em  The Hamiltonian~\eqref{MSall}, written in Beltrami coordinates,
is endowed with three integrals of motion quadratic in the momenta  and  given by
\bea 
   &&\ii_{1,\kk}= \frac{1}{2} \left( J_{01}^2+\kk J_{12}^2 \right)   + \del\, \frac{\tq_1^2(1+\kk \tq_2^2) }{(1-\kk \tq_2^2)^2}  +\la_1\,\frac{(1+\kk \btq^2)}{\tq_1^2}, \nonumber \\ 
     &&\ii_{2,\kk}=\frac{1}{2} J_{02}^2 +4\del\,   \frac{\tq_2^2}{(1-\kk \tq_2^2)^2 } 
   ,  \label{md}      \\[2pt]
     &&\ele_{\kk}= J_{01} J_{12}  +2\del\,   \frac{\tq_1^2\tq_2}{(1-\kk \tq_2^2)^2 }-2\la_1\,  \frac{\tq_2}{\tq_1^2} ,\nonumber
 \eea
where ${\cal T}_\k$,  $J_{\mu\nu}$ are provided in table 1 and   $\non_\kk=\ii_{1,\kk}+\ii_{2,\kk}$.  The two sets  $(\non_\kk , \ii_{1,\kk} ,\ele_{\kk}  )$ and $(\non_\kk, \ii_{2,\kk} , \ele_{\kk} )$ are formed by three functionally independent functions.
 }
\medskip


\sect{A new  integrable  curved  anisotropic   oscillator}

In~\cite{Non} the problem of the construction of an integrable anisotropic generalization of the Higgs oscillator~\eqref{Higgsall} was faced, and the following integrable Hamiltonian ${\cal H}_\k^{\del,\Om}$ was found, which can be rewritten in terms of the four types of canonical variables as:
  \bea
  &&{\cal H}_\k^{\del,\Om}= {\cal T}_\k+  \del \,\frac{\>x^2}{(1-\k \>x^2)} +\Om\, x_2^2 +\frac{\la_1}{x_1^2} +\frac{\la_2}{x_2^2} \nonumber\\
  && \quad\quad\  = {\cal T}_\k+\del \Tk^2_\kk(r)+\Om \Sk^2_\kk(r)\sin^2\te+\frac{\la_1}{\Sk^2_\kk(r) \cos^2\te}+\frac{\la_2}{\Sk^2_\kk(r) \sin^2\te} 
 \nonumber
  \\
   &&\quad\quad\  =  {\cal T}_\k+   \del\,\frac{4 \bttq^2}{(1-\k\bttq^2)^2} + \Om \,\frac{4 \ttq_2^2}{(1+\k\bttq^2)^2}  + \frac 14 {(1+\k\bttq^2)^2}   \left(  \frac{\la_1}{  \ttq_1^2} +\frac{\la_2} { \ttq_2^2}  \right)  \nonumber
  \\
   &&\quad\quad\  =  {\cal T}_\k+ \del \bq^2 +\Om\, \frac{q_2^2}{(1+\kk \bq^2)}+
\left(1+\kk \bq^2\right)\left(\frac{\la_1}{q_1^2}+\frac{\la_2}{q_2^2} \right).
    \label{Higgsallanis}
  \eea
 This `anisotropic Higgs oscillator' only differs from~\eqref{Higgsall} in the term containing $\Om$, and the  Higgs system is recovered in the isotropic limit $\Om\to 0$. The Hamiltonian ${\cal H}_\k^{\del,\Om}$ has one independent integral of the motion, which is quadratic in the momenta, and numerical integration for many different values of $\Om$ and initial conditions seems to indicate that this Hamiltonian system is superintegrable {\em only in the case $\Om=0$} (see \cite{Non} for a detailed discussion). 
 
Moreover, if we take $\Om=3\delta$ in~\eqref{Higgsallanis}, the anisotropic $1:2$ Hamiltonian~\eqref{MSall} {\em is not recovered}. Indeed, for the latter system all bounded trajectories are periodic (it is a superintegrable system), whilst for the former one that is not the case. Thus, both of them are completely different curved generalizations of the same Euclidean system with potential ${\cal U}_0^{\delta,3\delta} =\delta (\tq_1^2 + 4 \tq_2^2)$. 

Consequently, we conjectured in~\cite{Non} that {\em each} of the Euclidean superintegrable anisotropic  oscillators with commensurate frequencies~\cite{Jauch, Tempesta} would be the $\k\to 0$ limit of a different family of integrable curved anisotropic oscillators. In particular, the first step in order to support this conjecture would consist in obtaining a new curved anisotropic oscillator that, being integrable for any value of the anisotropy parameter, would reduce to the superintegrable $1:2$ oscillator when the anisotropy parameter takes the appropriate values. 

Such a new integrable Hamiltonian system constitutes the main result of this paper, that is summarized in the following stament.

\medskip

\noindent
{\bf Proposition 2.}  {\em  Let  $\nonn_\kk$ be the following Hamiltonian     written in Beltrami variables   
\be
\nonn_\kk= {\cal T}_\k+\Om_1\,\frac{  \tq_1^2}{(1+\kk  \tq_2^2)}
+\Om_2\,\frac{  (1+\kk \btq^2)  \tq_2^2}{ (1+\kk  \tq_2^2)(1-\kk  \tq_2^2)^2  } +
\left(1+\kk \bq^2\right)\left(\frac{\la_1}{q_1^2}+\frac{\la_2}{q_2^2} \right).
 \label{na}
    \ee
Then, for any value of the real constants $\{\k,\Om_1,\Om_2,\lambda_1,\lambda_2\}$,  the Hamiltonian system $\nonn_\kk$ is integrable, and  its integrals of motion (that are quadratic in the momenta) are given by
\bea 
   &&\!\!\!  \!\!\!\!\!\!\!\!\!   \iii_{1,\kk}= \frac{1}{2} \left( J_{01}^2+\kk J_{12}^2 \right)   + \Om_1\, \frac{\tq_1^2(1+\kk \tq_2^2) }{(1-\kk \tq_2^2)^2}
   +\k\left(\Om_2-4\Om_1 \right) \frac{\tq_1^2\tq_2^2}{ (1+\kk  \tq_2^2)(1-\kk  \tq_2^2)^2  } 
\nonumber      \\
   &&\qquad\qquad    +\la_1\,\frac{(1+\kk \btq^2)}{\tq_1^2} +\la_2\k \,\frac{ \btq^2}{\tq_2^2}, \nonumber \\ 
     &&\!\!\!  \!\!\!\!\!\!\!\!\!   \iii_{2,\kk}=\frac{1}{2} J_{02}^2 +\Om_2\,   \frac{\tq_2^2}{(1-\kk \tq_2^2)^2 }  + \frac{\la_2}{\tq_2^2},
\nonumber
      \eea
and are such that $\nonn_\kk=\iii_{1,\kk}+\iii_{2,\kk}$, where ${\cal T}_\k$ and  $J_{\mu\nu}$ are the functions given in table 1.}
\medskip

This result  can be proven through straightforward computations. Several comments are in order:

\begin{itemize}

\item When $\Om_1=\delta$,  $\Om_2=4\delta$ and $\lambda_2=0$, the maximally superintegrable $1:2 $ oscillator given in Proposition 1 is recovered together with two of its (non-independent) integrals of the motion, although the third independent one $\ele_{\kk}$ ensuring the superintegrability of the system cannot be obtained from Proposition 2. Therefore, the system~\eqref{na} can be properly  called the `anisotropic generalization' of the system~\eqref{MSall}.

\item However, when $\Om_1=\Om_2$ the Hamiltonian (\ref{na}) is by no means the  Higgs oscillator (\ref{Higgsall}) with $\la_2=0$. Indeed, $\nonn_\kk$ does not coincide with the anisotropic Higgs oscillator~\eqref{Higgsallanis} either.  Therefore, for arbitrary values of $\Om_1$ and $\Om_2$  the system $\nonn_\kk$ defines a new integrable  curved  generalization of the Euclidean aniso\-tropic oscillator~\eqref{aa}. 

\item The Rosochatius terms containing $\la_1$ and $\la_2$ potentials are proper {\em centrifugal barriers} when both  $\lambda_1$ and $\lambda_2$ are {\em positive}. In that case these terms can be interpreted (see~\cite{Non} for a complete geometric discussion) as noncentral  oscillators on $\>S^2$ ($\k>0$) with centres (in ambient coordinates) located at $O_1=(0,1,0)$ and $O_2=(0,0,1)$, respectively.
Recall that the Higgs oscillator (\ref{Higgsall})  is a central oscillator with centre at the origin $O=(1,0,0)$ for any value of $\k$.

\end{itemize}

By making use of table  1, Proposition 2 can be indeed rewritten in terms of ambient, polar and Poincar\'e variables.
In particular, the potential ${\cal U}_\k$ of (\ref{na}) reads:
  \bea
  &&{\cal U}_\k=\Om_1\,\frac{x_1^2}{(1-\k x_1^2)}+\Om_2\,\frac{x_0^2 x_2^2}{ (x_0^2+\k x_2^2)(x_0^2-\k x_2^2)^2} +\frac{\lambda_1}{x_1^2} +\frac{\lambda_2}{x_2^2} \nonumber\\
  && \quad\  =\Om_1 \,\frac{\Sk^2_\kk(r)\cos^2\te}{\left(1-\kk \Sk^2_\kk(r)\cos^2\te\right)}     +\Om_2\, \frac{\Tk^2_\kk(r)\sin^2\te}{\left(1-\kk \Sk^2_\kk(r)\cos^2\te\right)\left(1-\kk \Tk^2_\kk(r)\sin^2\te\right)^2} \nonumber
  \\
  &&\qquad \qquad +\frac{\la_1}{\Sk^2_\kk(r) \cos^2\te}+\frac{\la_2}{\Sk^2_\kk(r) \sin^2\te} \nonumber \\
  &&\quad\  =  \Om_1\,\frac{4 \ttq_1^2}{  (1+\k\bttq^2)^2-4\k \ttq_1^2} + \Om_2\,\frac{4\ttq_2^2(1+\k\bttq^2)^2(1-\k\bttq^2)^2}{    \left( (1-\k\bttq^2)^2+4\k \ttq_2^2 \right ) \left( (1-\k\bttq^2)^2-4\k \ttq_2^2 \right )^2  }  \nonumber \\
  &&\qquad \qquad + \frac 14 {(1+\k\bttq^2)^2}   \left(  \frac{\la_1}{  \ttq_1^2} +\frac{\la_2} { \ttq_2^2}  \right)  .
  \label{nc}
  \eea
All these expressions allows us to confirm that, to the best of our knowledge, this potential differs from other curved anisotropic oscillators given in the literature~\cite{Non, Nerssesian,kalnins, Saksida}. On the other hand, it becomes evident that --again-- Beltrami variables are the simplest ones in order to deal with this kind of systems.


\begin{figure}[t]
\setlength{\unitlength}{1mm}
\begin{picture}(140,132)(0,0)
\label{figure1}
\footnotesize{
\put(11,70){\includegraphics[scale=0.25]{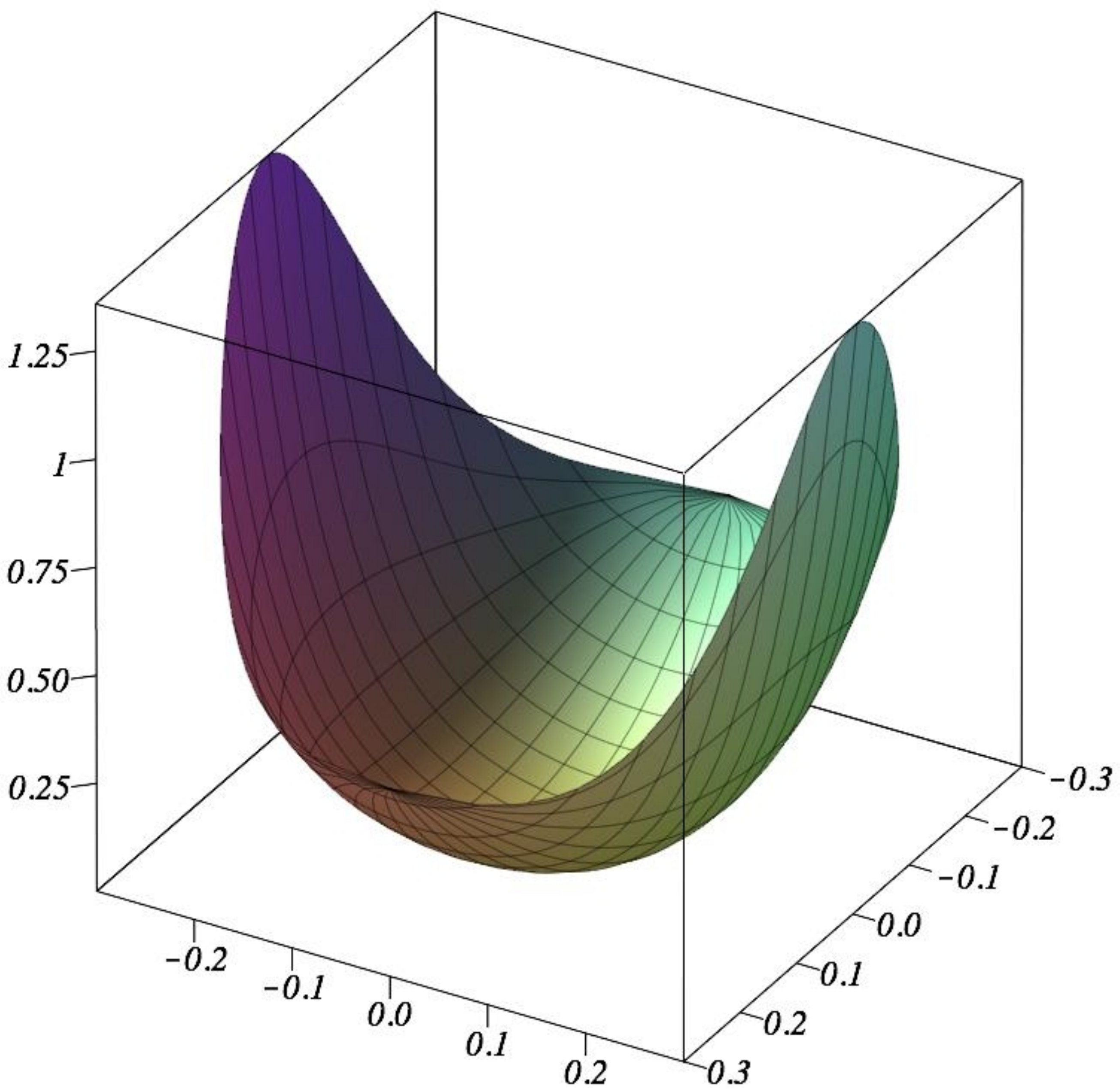}}
\put(28,72){\footnotesize $\ttq_2$}
\put(70,78){\footnotesize $\ttq_1$}
\put(10,122){\footnotesize ${\cal U}_+$}
\put(11,72){(a)}
\put(85,70){\includegraphics[scale=0.25]{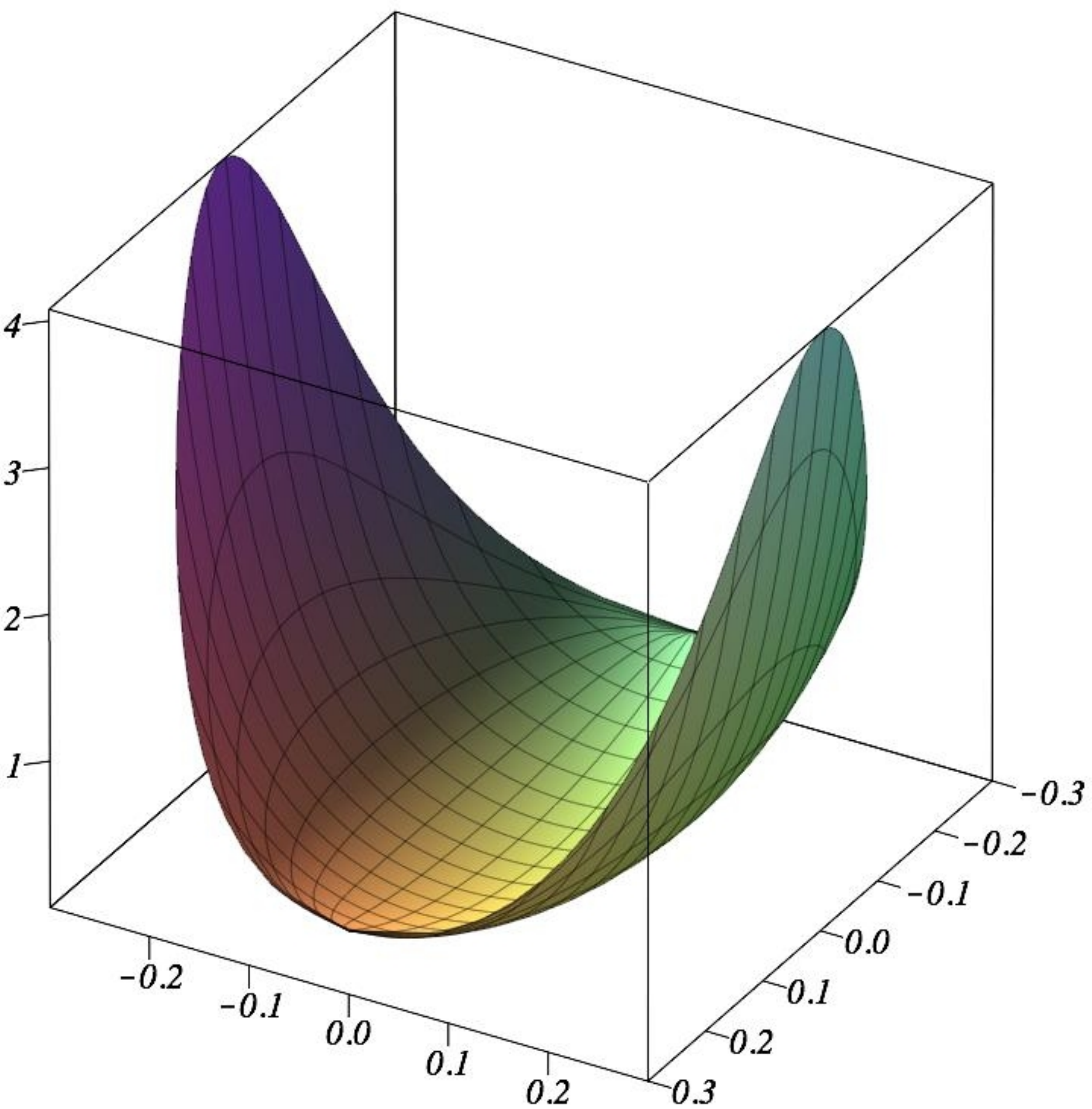}}
\put(102,72){\footnotesize $\ttq_2$}
\put(142,78){\footnotesize $\ttq_1$}
\put(84,122){\footnotesize ${\cal U}_+$}
\put(85,72){(b)}
\put(11,2){\includegraphics[scale=0.25]{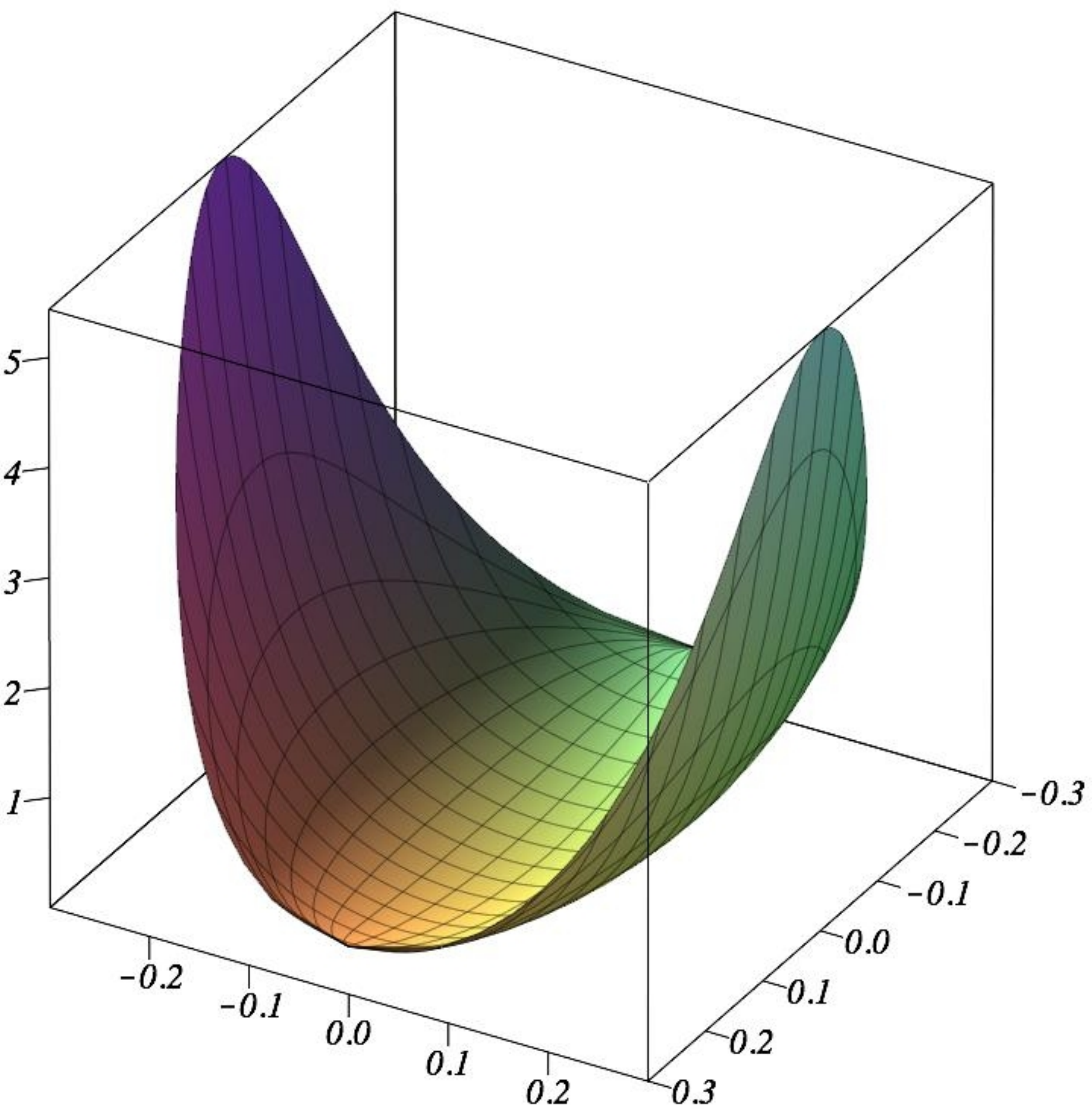}}
\put(28,3){\footnotesize $\ttq_2$}
\put(67,9){\footnotesize $\ttq_1$}
\put(10,53){\footnotesize ${\cal U}_+$}
\put(11,3){(c)}
\put(85,2){\includegraphics[scale=0.25]{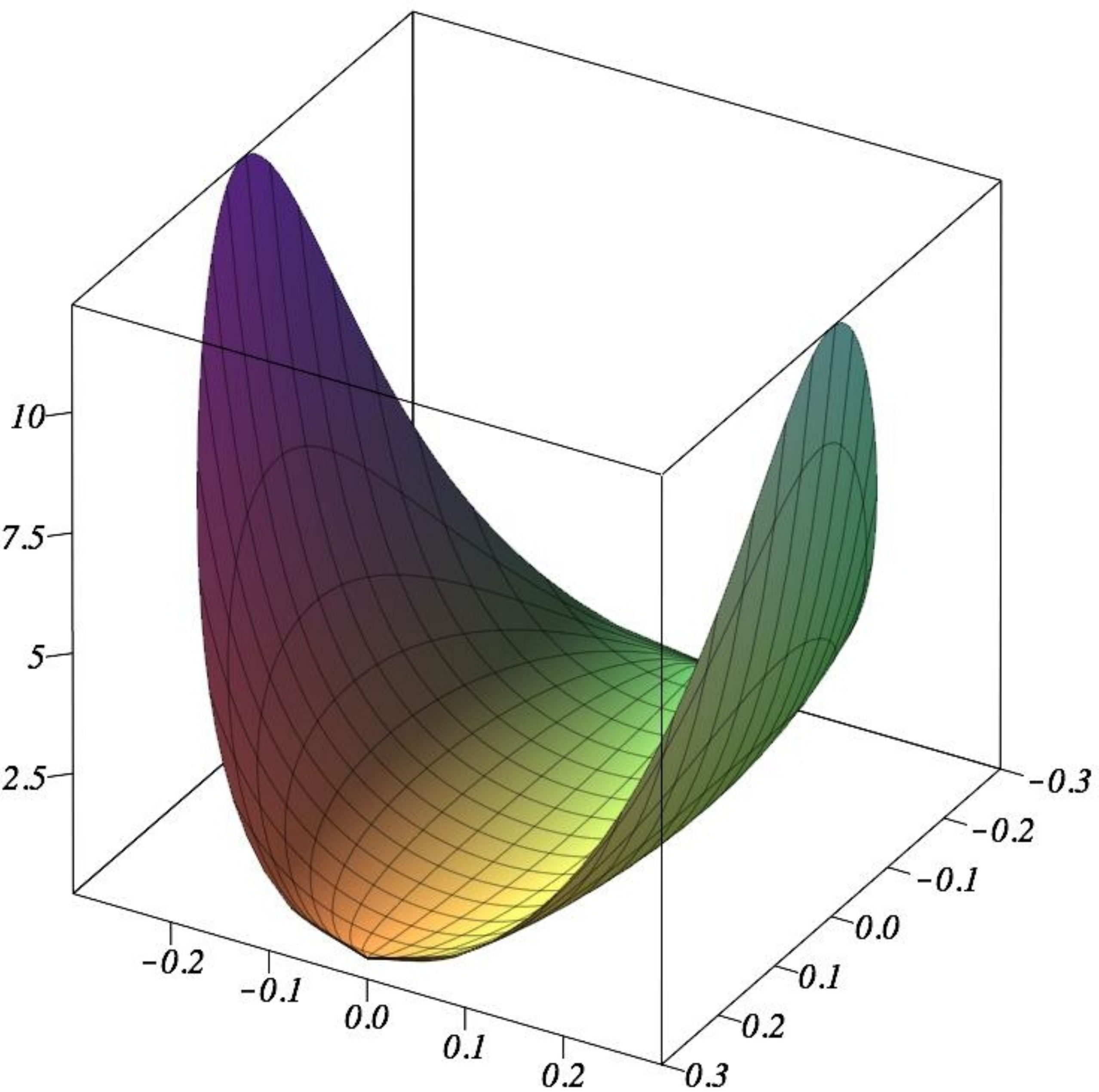}}
\put(102,3){\footnotesize $\ttq_2$}
\put(142,9){\footnotesize $\ttq_1$}
\put(84,53){\footnotesize ${\cal U}_+$}
\put(85,3){(d)}
}
\end{picture}
\caption{ \footnotesize The potential ${\cal U}_\k$ (\ref{nc}) on the sphere ${\mathbf S}^2$ in Poincar\'e coordinates $(\ttq_1,\ttq_2)$, with $\k=+1$,  $\Om_1=1$ and without Rosochatius terms  $\la_1=\la_2=0$:
 (a) $\Omega_2=1$ (the     $1:1$  case),
(b)     $\Omega_2=3$,
(c) $\Omega_2=4$  (the superintegrable  $1:2$  case),
 and 
(d)  $\Omega_2=9$ (the     $1:3$  case). No essential differences are appreciated in the superintegrable case~(c).}
\end{figure}
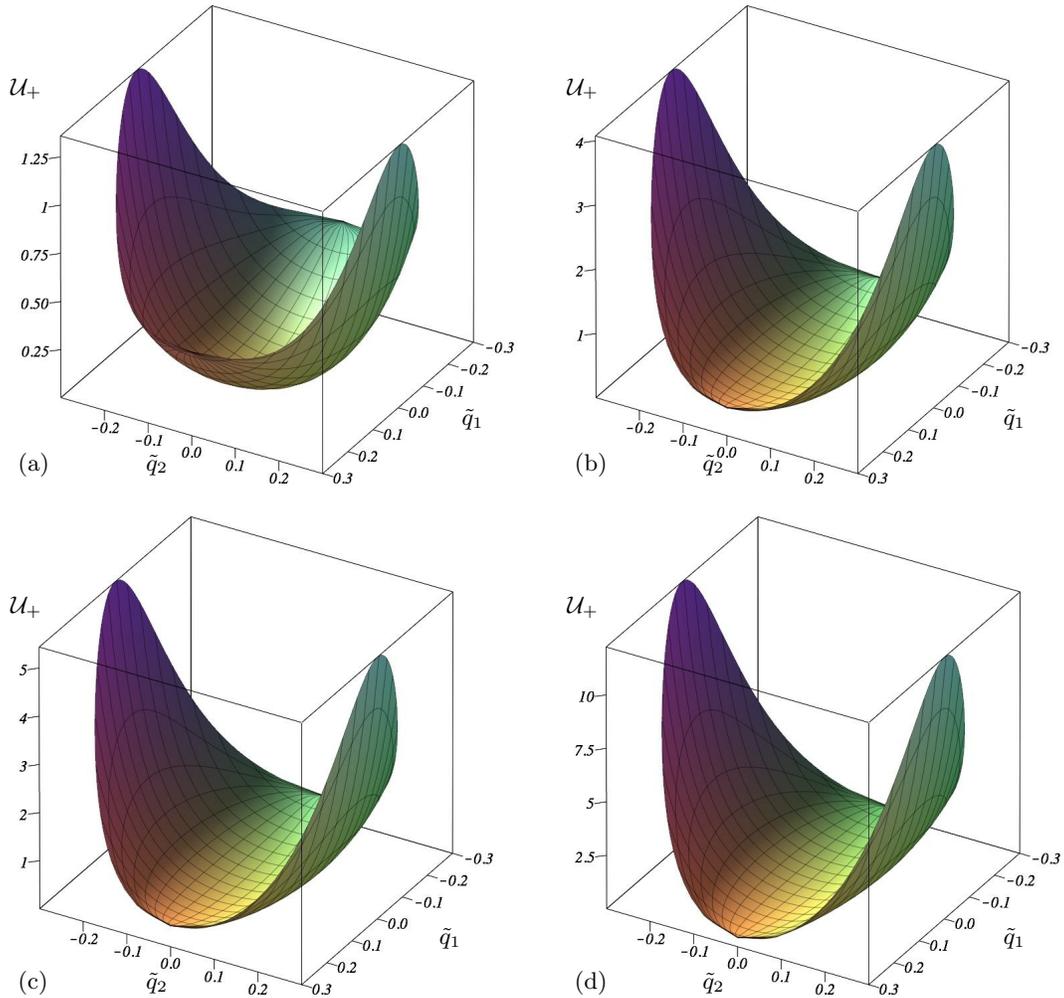



\begin{figure}[t]
\setlength{\unitlength}{1mm}
\begin{picture}(140,132)(0,0)
\label{figure2}
\footnotesize{
\put(11,69){\includegraphics[scale=0.25]{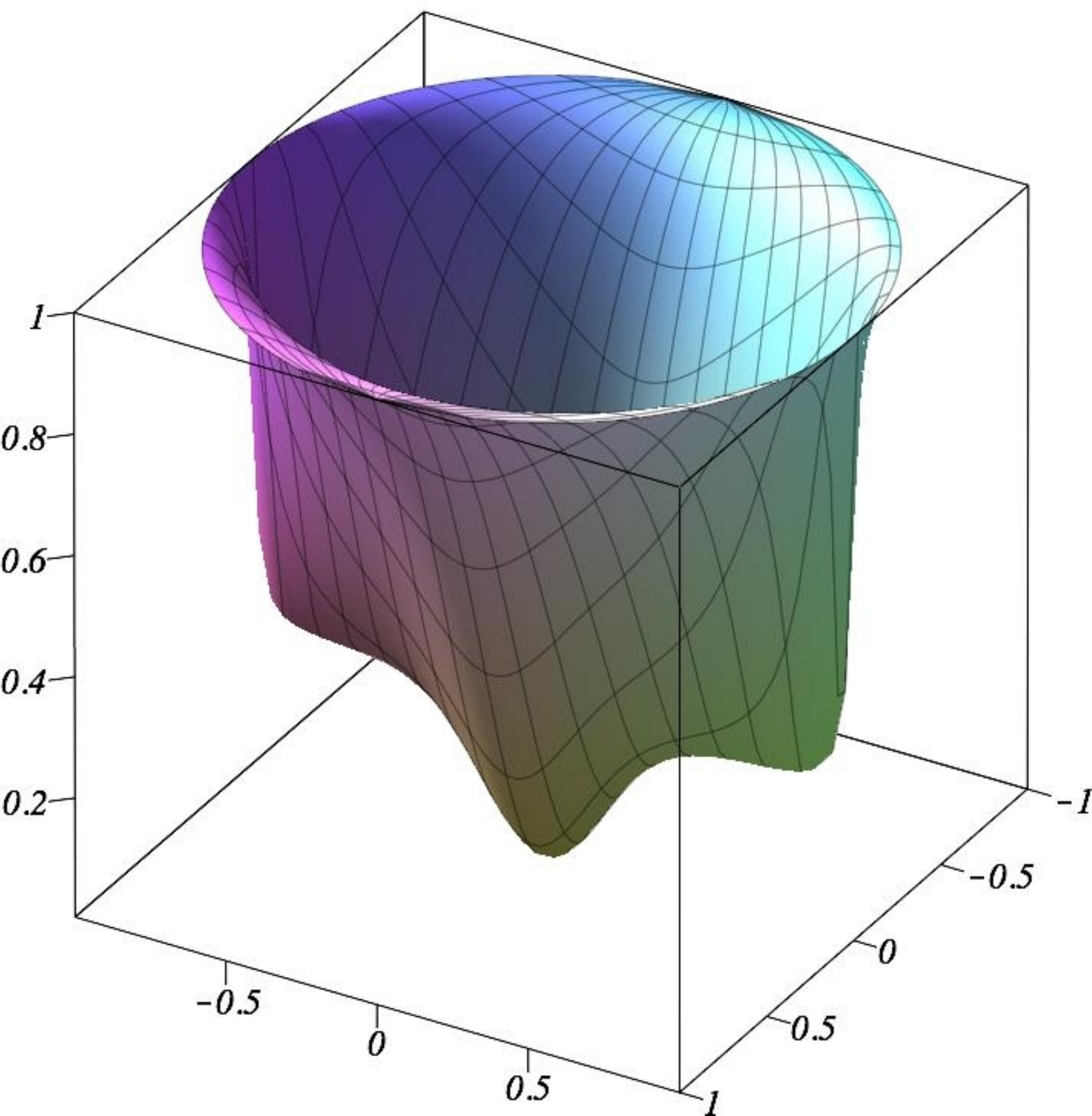}}
\put(28,72){\footnotesize $\ttq_2$}
\put(68,78){\footnotesize $\ttq_1$}
\put(10,122){\footnotesize ${\cal U}_-$}
\put(11,72){(a)}
\put(85,69){\includegraphics[scale=0.25]{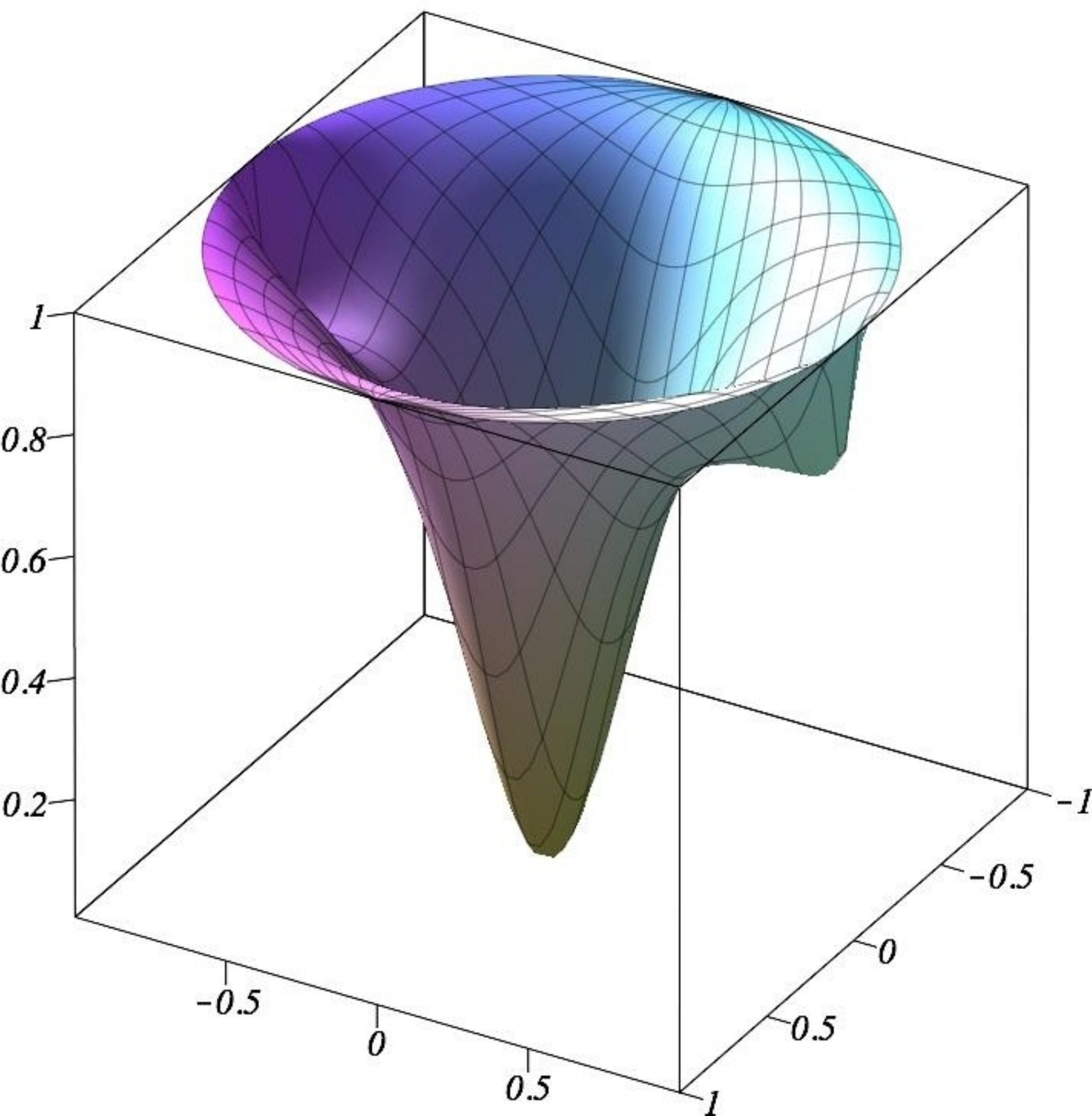}}
\put(102,72){\footnotesize $\ttq_2$}
\put(142,78){\footnotesize $\ttq_1$}
\put(84,122){\footnotesize ${\cal U}_-$}
\put(85,72){(b)}
\put(11,0){\includegraphics[scale=0.25]{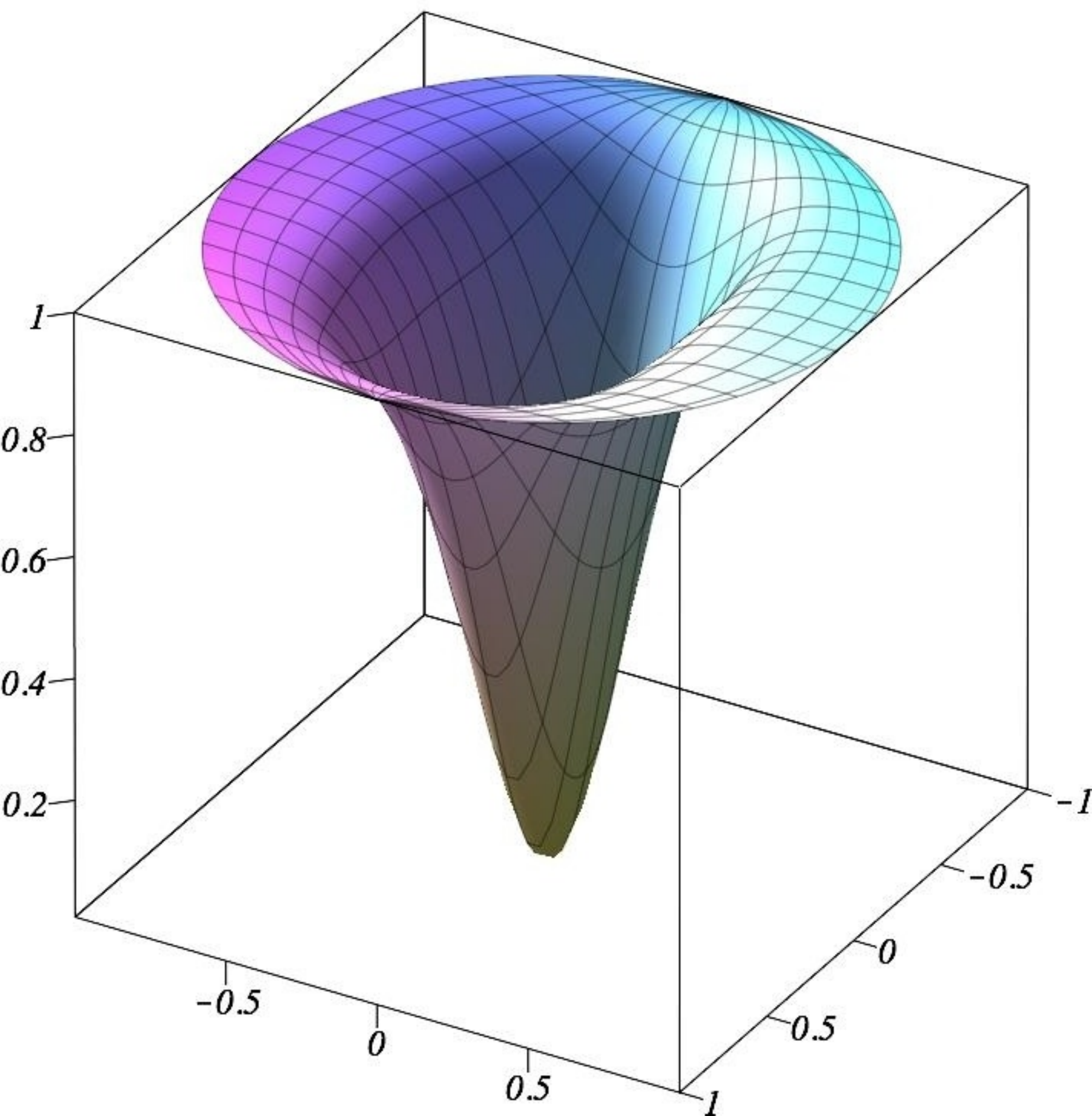}}
\put(28,3){\footnotesize $\ttq_2$}
\put(68,9){\footnotesize $\ttq_1$}
\put(10,53){\footnotesize ${\cal U}_-$}
\put(11,3){(c)}
\put(85,0){\includegraphics[scale=0.25]{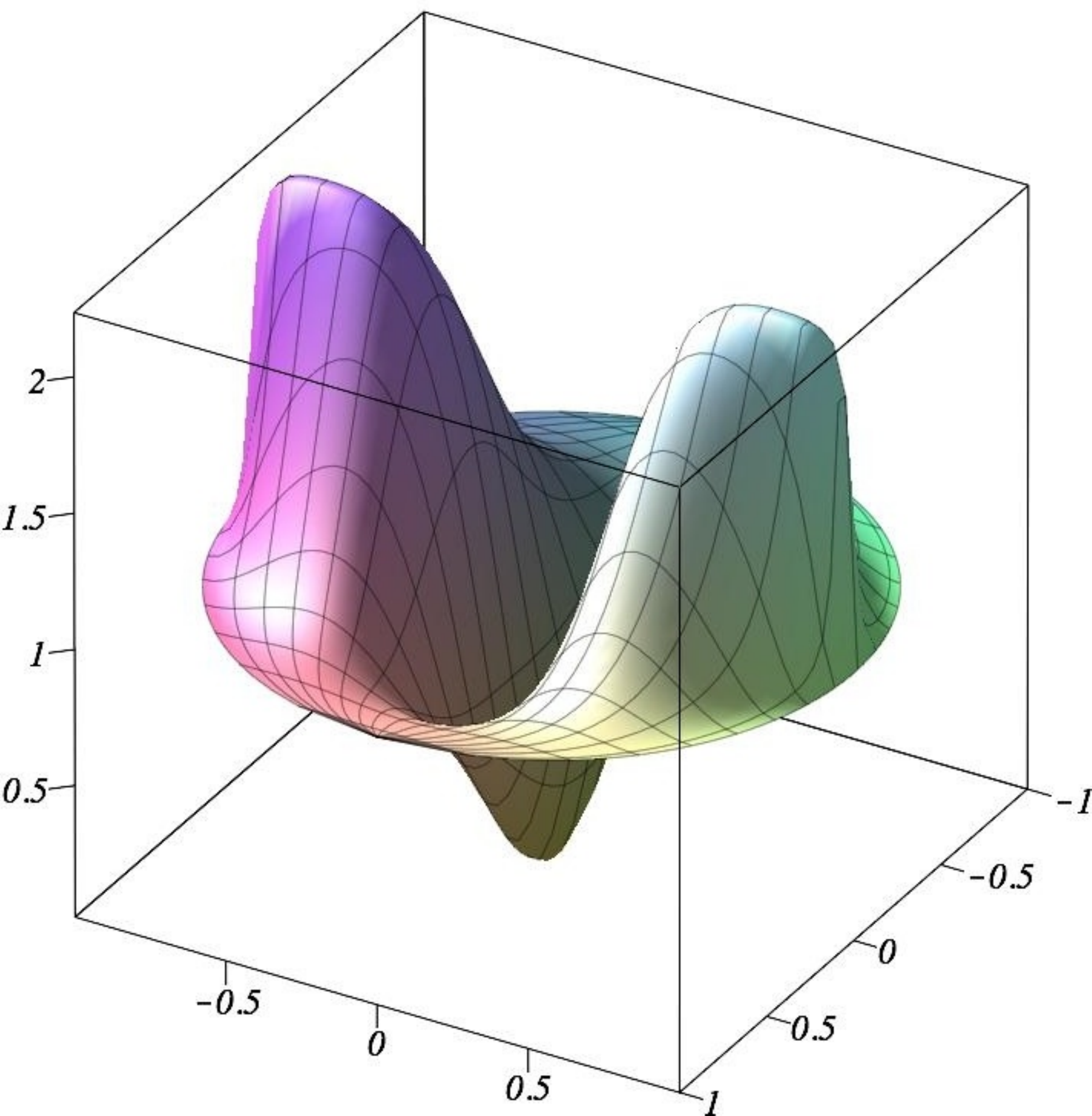}}
\put(102,3){\footnotesize $\ttq_2$}
\put(142,9){\footnotesize $\ttq_1$}
\put(84,53){\footnotesize ${\cal U}_-$}
\put(85,3){(d)}
}
\end{picture}
\caption{ \footnotesize The potential ${\cal U}_\k$ (\ref{nc})  on the hyperboloid ${\mathbf H}^2$ in Poincar\'e coordinates $(\ttq_1,\ttq_2)$  with $\k=-1$,  $\Om_1=1$ and $\la_1=\la_2=0$:
 (a) $\Omega_2=1$ (the     $1:1$  case),
(b)     $\Omega_2=3$,
(c) $\Omega_2=4$  (the superintegrable  $1:2$  case),
 and 
(d)  $\Omega_2=9$  (the     $1:3$  case). It turns out that, when plotted on the Poincar\'e disk, the geometric shape of the superintegrable case (c) is quite different from the other (integrable) potentials.}
\end{figure}


However, although the expression for the potential in terms of Poincar\'e variables (\ref{nc})  is quite cumbersome, these projective coordinates present an unexpected and interesting feature from the integrability viewpoint: if the potential ${\cal U}_\k$ for the hyperbolic space is represented  on the Poincar\'e disk  (figure 2), the superintegrable case $\Om_1=1, \Om_2=4$ is neatly distinguished from any other value of the anisotropy parameters, while the same superintegrable potential on the sphere does not present any singular feature (figure 1). However, the same plot in Beltrami coordinates does not provide such a visual approach to the superintegrability properties of the potential (recall that the Beltrami coordinates do not cover the complete sphere as the proyection of the points on the equator $x_0=0$ goes to infinity). These facts seem to indicate that the use of projective phase spaces can be meaningful from a qualitative viewpoint and furthermore, as we will see in the sequel, for the numerical integration of the dynamics.


\sect{Numerical integration and superintegrability}

So far, we have obtained a new family of  anisotropic curved oscillators $\nonn_\kk$~\eqref{na} that generalize the curved superintegrable $1:2$ system~\eqref{MSall}.  We have also explicitly proven the complete integrability of $\nonn_\kk$, but its superintegrability is only ensured for  the particular case with $\Om_2=4\Om_1 $ and $\la_2=0$, which is just  (\ref{MSall}). Due to \cite{RS}, we know that this is the only possible superintegrable case of $\nonn_\kk$ whose constants of motion are {\em quadratic} in the momenta.     

However, this poses the question of whether there exist other values of $\Om_1$ and $\Om_2$ for which $\nonn_\kk$ is also superintegrable. If that would be the case, the additional integral of the motion (that would play the role of $\ele_{\kk}$~\eqref{md} for the  $1:2$ case) should be of higher degree in the momenta and it should depend analytically on the curvature parameter $\k$. Moreover, since the Euclidean limit $\k\to 0$ is always well defined, the only possibilities for the superintegrability of $\nonn_\kk$ would be provided by the choices for the  $\Om_1$ and $\Om_2$ parameters that lead to superintegrable systems in the Euclidean limit, {\em i.e.}, for values $(\Om_1,\Om_2)$ that give rise to commensurate frequencies when $\k=0$.


\begin{figure}[t]
\setlength{\unitlength}{1mm}
\begin{picture}(140,132)(0,0)
\label{figure3}
\footnotesize{
\put(11,69){\includegraphics[scale=0.25]{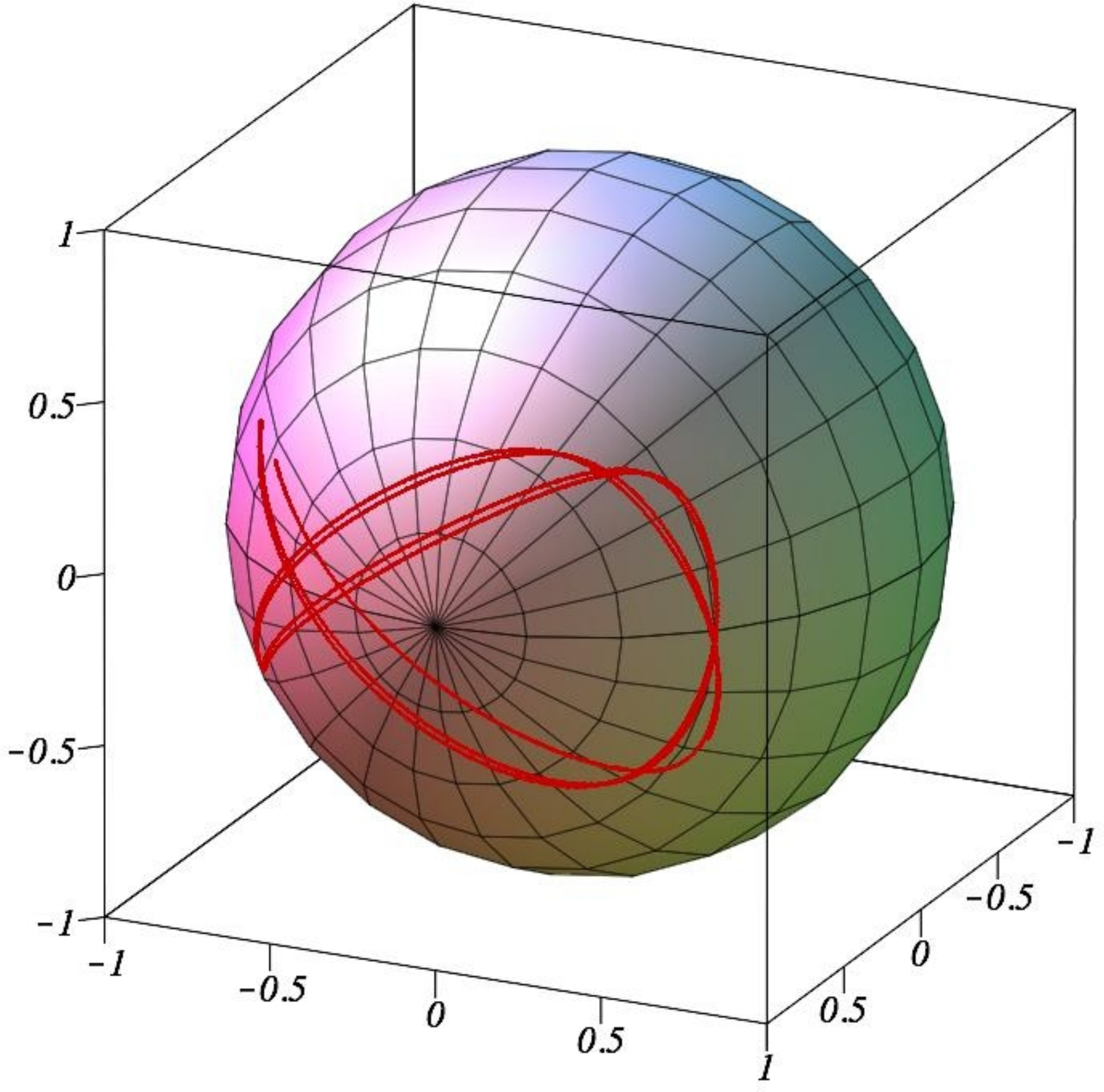}}
\put(30,70){\footnotesize $x_1$}
\put(68,75){\footnotesize $x_0$}
\put(12,121){\footnotesize $x_2$}
\put(11,72){(a)}
\put(85,69){\includegraphics[scale= 0.25]{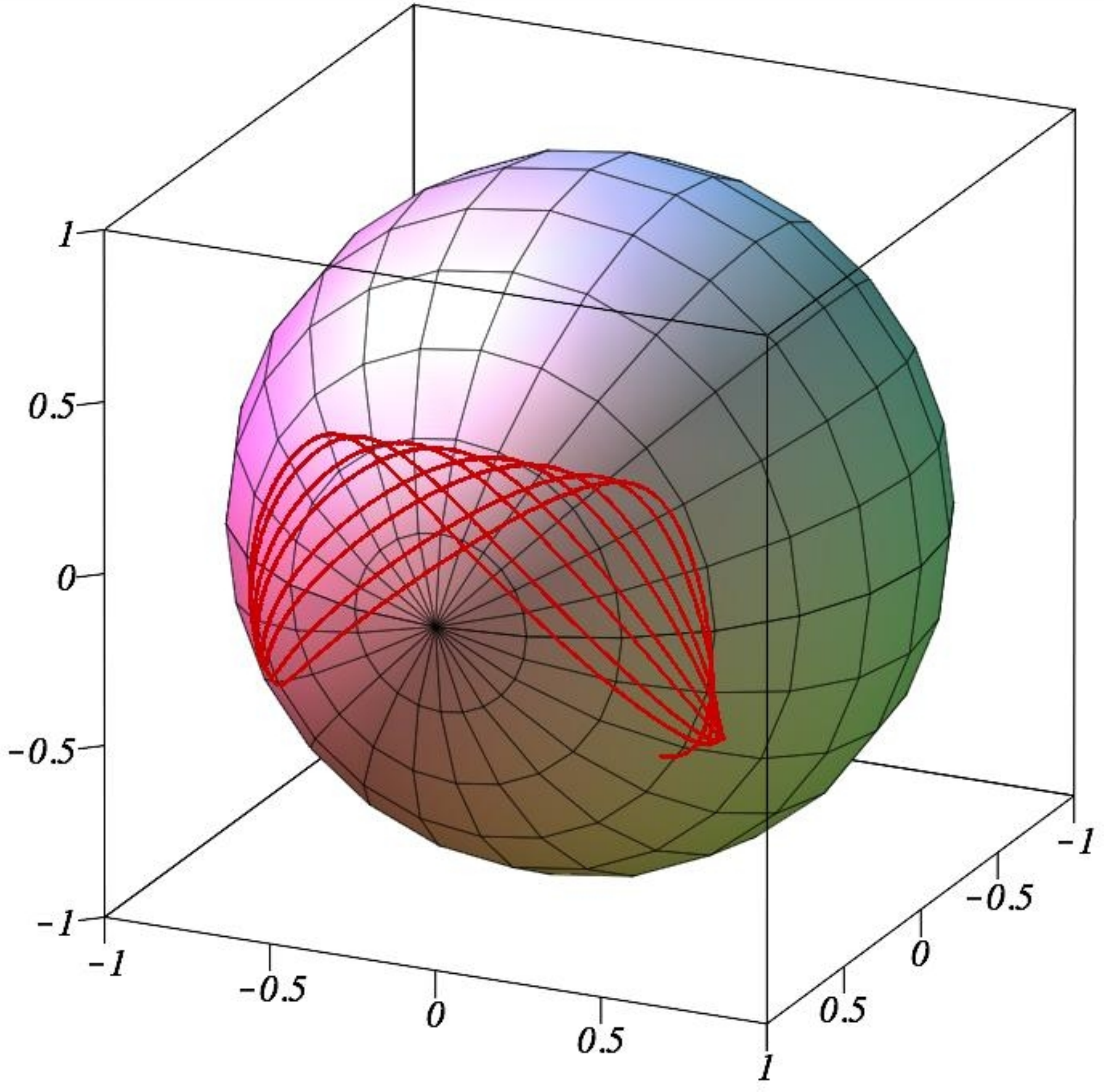}}
\put(104,70){\footnotesize $x_1$}
\put(142,75){\footnotesize $x_0$}
\put(86,121){\footnotesize $x_2$}
\put(85,72){(b)}
\put(11,0){\includegraphics[scale= 0.25]{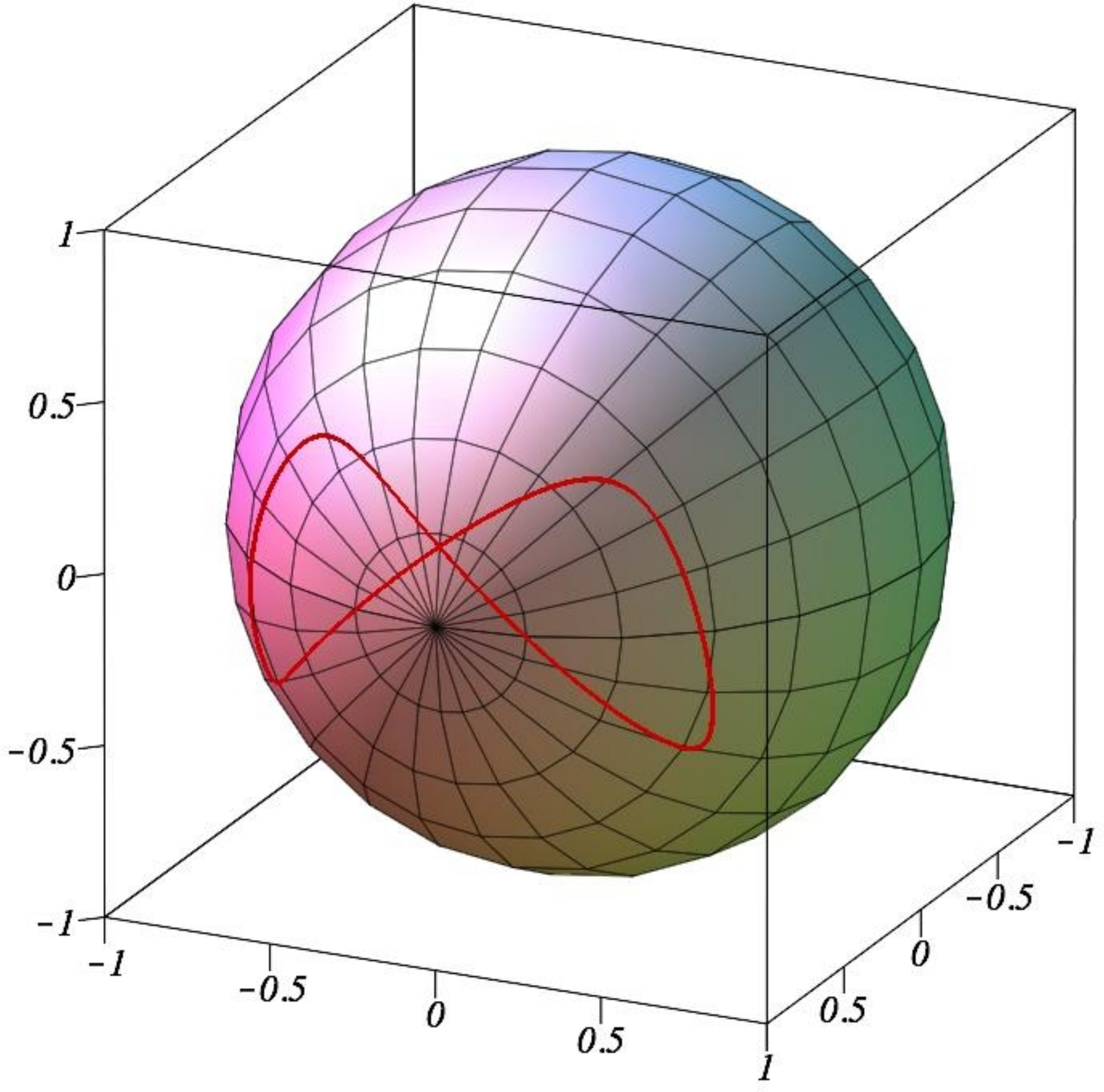}}
\put(30,1){\footnotesize $x_1$}
\put(68,6){\footnotesize $x_0$}
\put(12,52){\footnotesize $x_2$}
\put(11,3){(c)}
\put(85,0){\includegraphics[scale= 0.25]{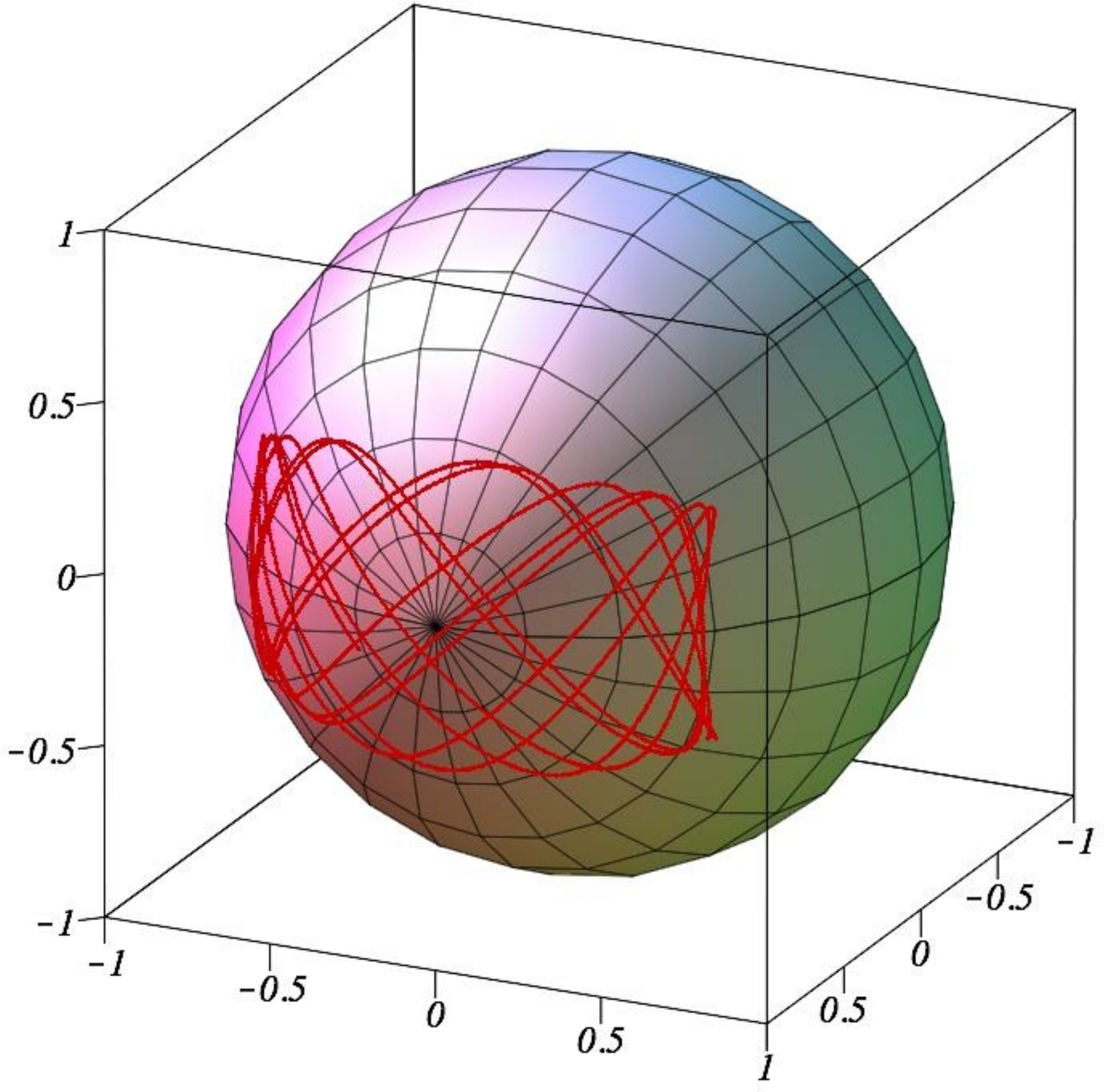}}
\put(104,1){\footnotesize $x_1$}
\put(142,6){\footnotesize $x_0$}
\put(86,52){\footnotesize $x_2$}
\put(85,3){(d)}
}
\end{picture}
\caption{ \footnotesize Some trajectories on the sphere  ${\mathbf S}^2$ for the Hamiltonian ${\cal H}_\kk$ (\ref{na})  
    with $\k=+1$, $\Om_1=1$ and without Rosochatius terms  $\la_1=\la_2=0$. They are plotted in  $\mathbb R^{3}$ with   ambient coordinates  $(x_0,\>x)$ fulfilling $x_0^2+x_1^2+x_2^2=1$. Time runs from $t=0$ to $t=8$ and the Hamilton equations are solved in terms of   Beltrami variables for the initial data  $\tq_1=1$, $\tq_2=-0.5$, $\dot{\tq}_1=1$, $\dot{\tq}_2=2$:
  (a) $\Omega_2=1$ (the    $ 1:1$  case),
(b)     $\Omega_2=3$,
(c) $\Omega_2=4$  (the superintegrable  $1:2$  case),
 and 
(d)  $\Omega_2=9$  (the    $ 1:3$  case). The only closed Lissajous-type trajectories correspond to the superintegrable case (c).}
\end{figure}



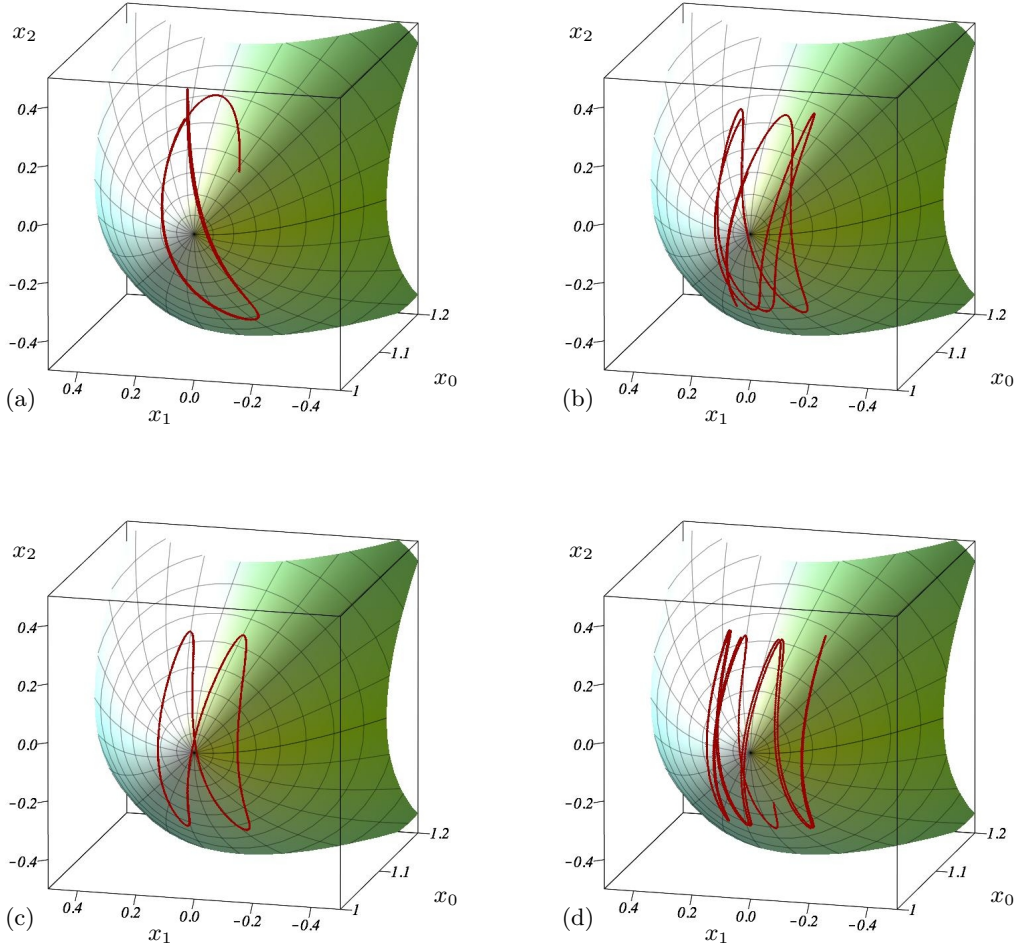
\begin{figure}[t]
\setlength{\unitlength}{1mm}
\begin{picture}(140,130)(0,0)
\label{figure4}
\footnotesize{
\put(11,71){\includegraphics[scale= 0.25]{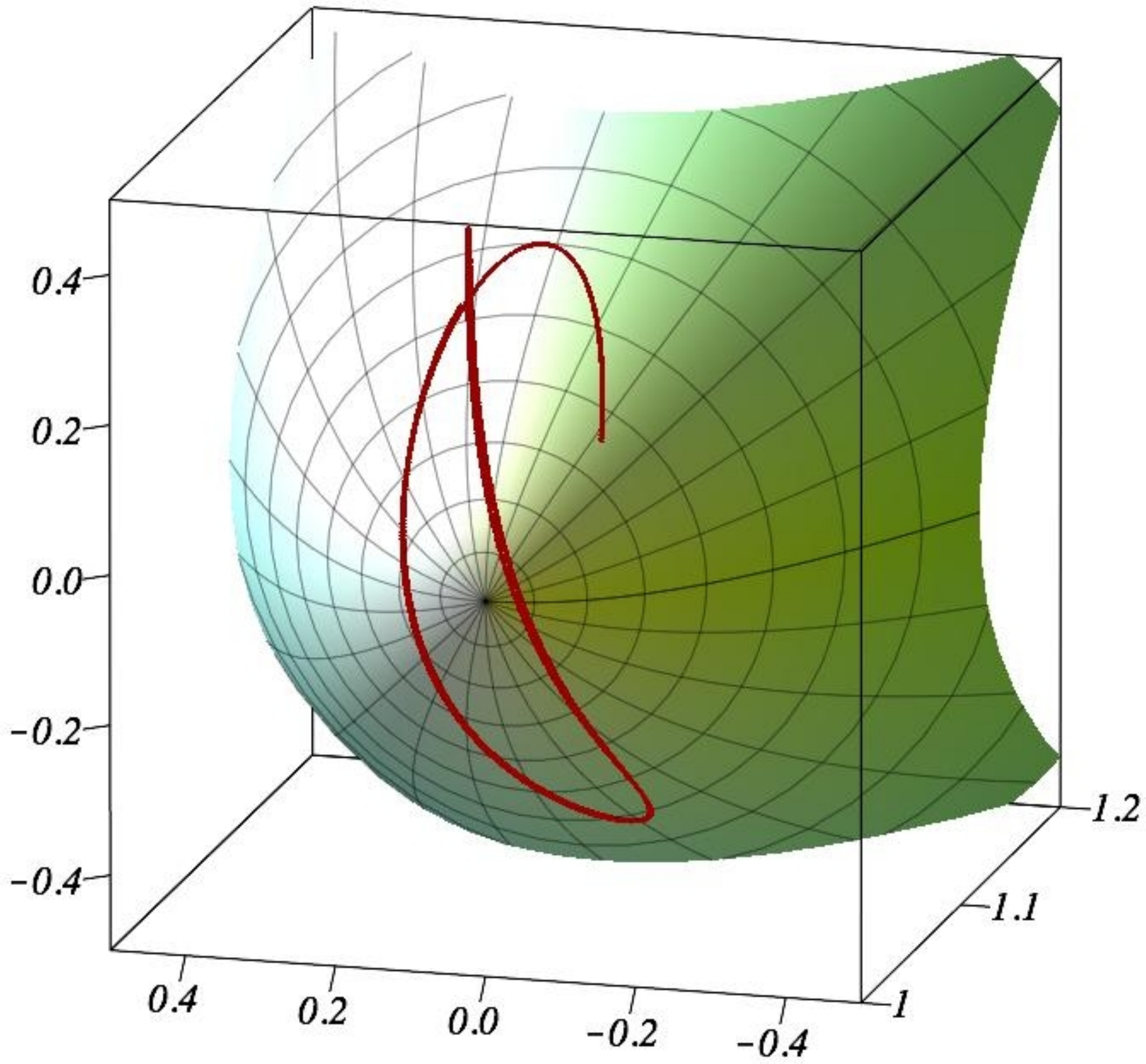}}
\put(30,70){\footnotesize $x_1$}
\put(68,75){\footnotesize $x_0$}
\put(12,121){\footnotesize $x_2$}
\put(11,72){(a)}
\put(85,71){\includegraphics[scale= 0.25]{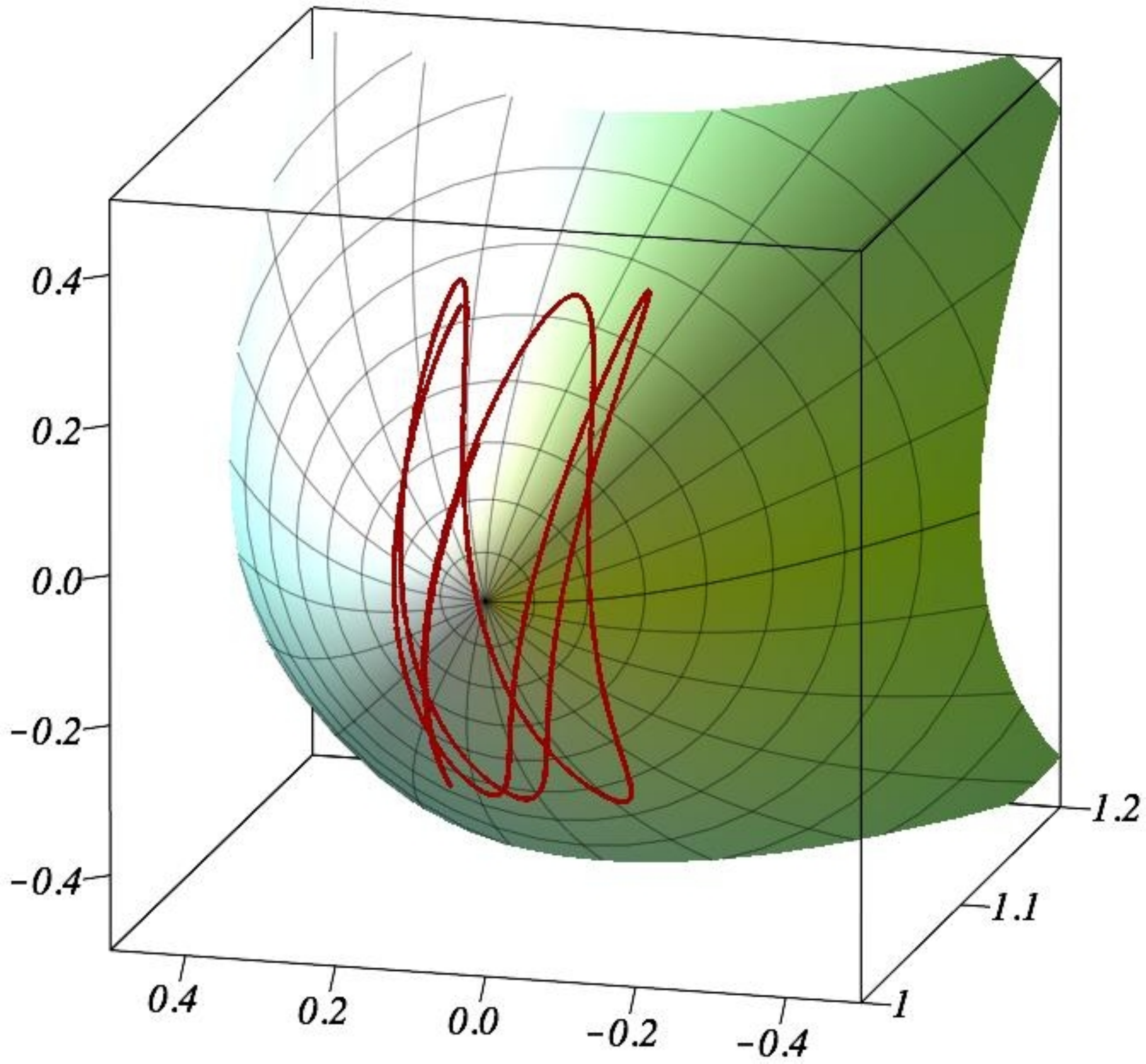}}
\put(104,70){\footnotesize $x_1$}
\put(142,75){\footnotesize $x_0$}
\put(86,121){\footnotesize $x_2$}
\put(85,72){(b)}
\put(11,2){\includegraphics[scale= 0.25]{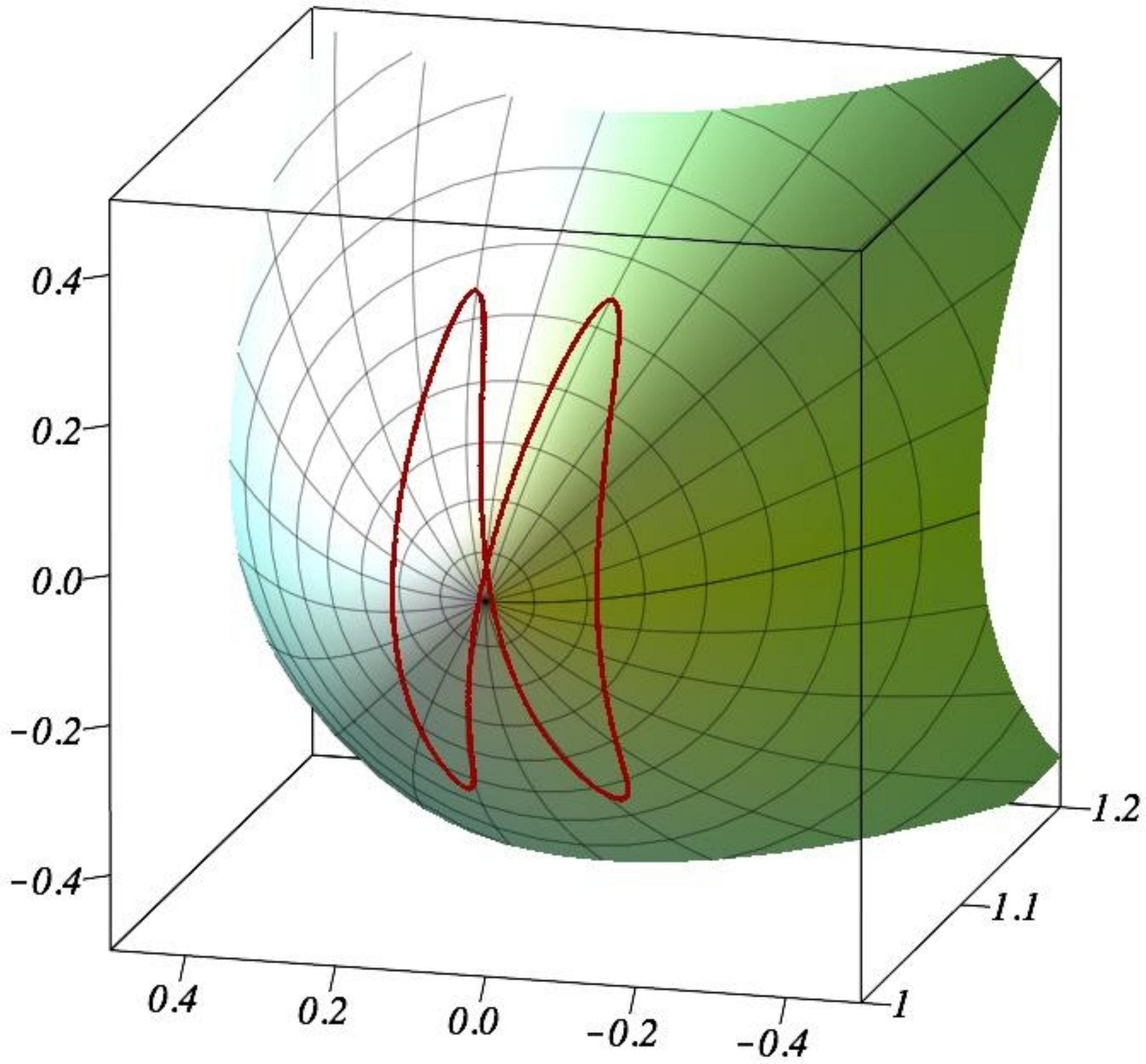}}
\put(30,1){\footnotesize $x_1$}
\put(68,6){\footnotesize $x_0$}
\put(12,52){\footnotesize $x_2$}
\put(11,3){(c)}
\put(85,2){\includegraphics[scale= 0.25]{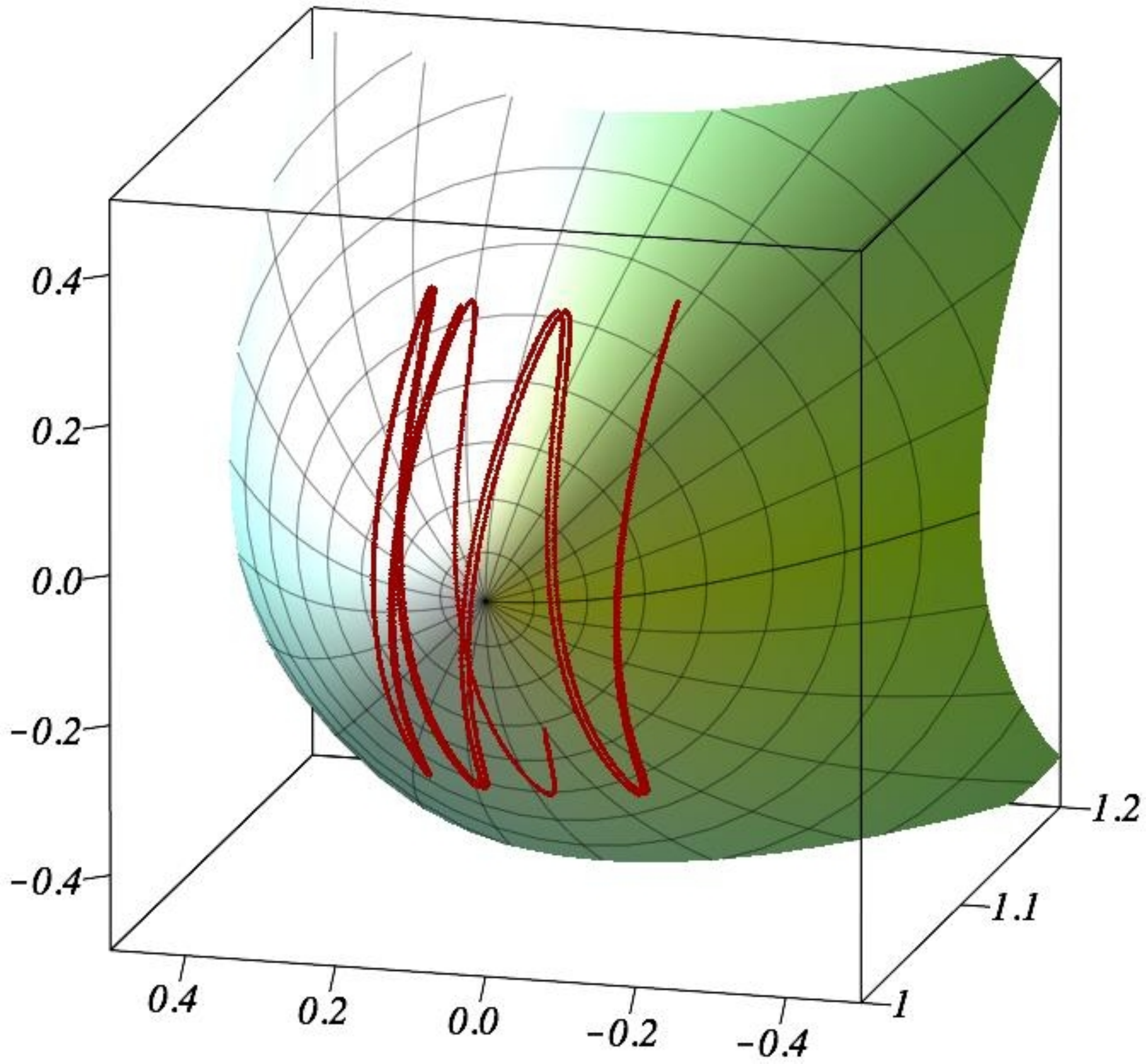}}
\put(104,1){\footnotesize $x_1$}
\put(142,6){\footnotesize $x_0$}
\put(86,52){\footnotesize $x_2$}
\put(85,3){(d)}
}
\end{picture}
\caption{ \footnotesize Some trajectories on the hyperboloid  ${\mathbf H}^2$ for the Hamiltonian ${\cal H}_\kk$ (\ref{na})  
      with $\k=-1$, $\Om_1=1$ and $\la_1=\la_2=0$. They are plotted in  $\mathbb R^{3}$ with   ambient coordinates  $(x_0,\>x)$ such that  $x_0^2-x_1^2-x_2^2=1$. Time runs from $t=0$ to $t=12$ and the numerical integration is again performed in Beltrami variables for the initial data $\tq_1=0.1$, $\tq_2=0.3$, $\dot{\tq}_1=0.1$, $\dot{\tq}_2=-0.2$:
  (a) $\Omega_2=1$ (the     $1:1$  case),
(b)     $\Omega_2=3$,
(c) $\Omega_2=4$  (the superintegrable  $1:2$  case),
 and 
(d)  $\Omega_2=9$  (the     $1:3$  case). Again, the superintegrable potential (c) is the only one that provides closed trajectories.}
\end{figure}



\begin{figure}[t]
\setlength{\unitlength}{1mm}
\begin{picture}(140,135)(0,0)
\label{figure5}
\footnotesize{
\put(11,69){\includegraphics[scale=0.3]{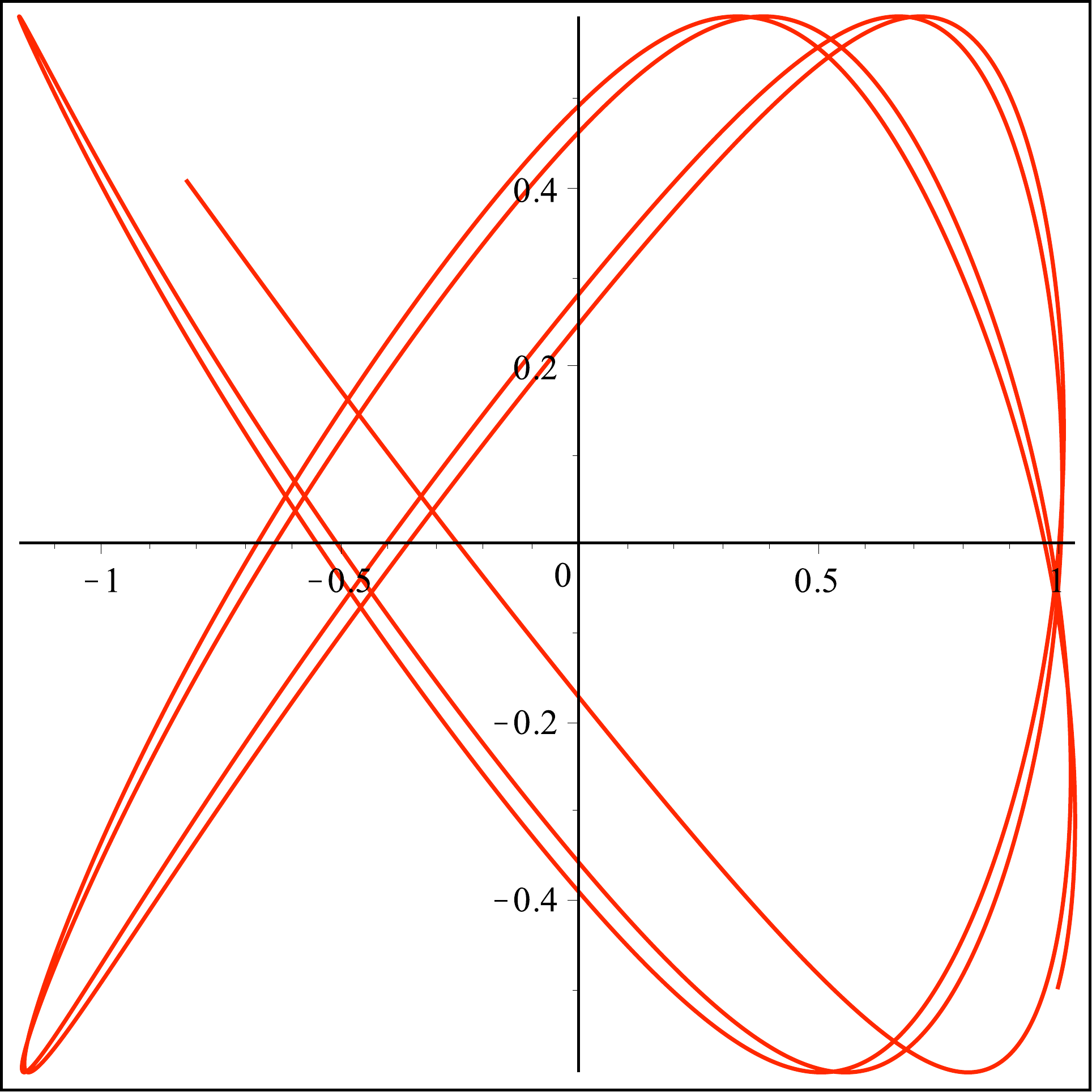}}
\put(5,72){(a)}
\put(85,69){\includegraphics[scale=0.3]{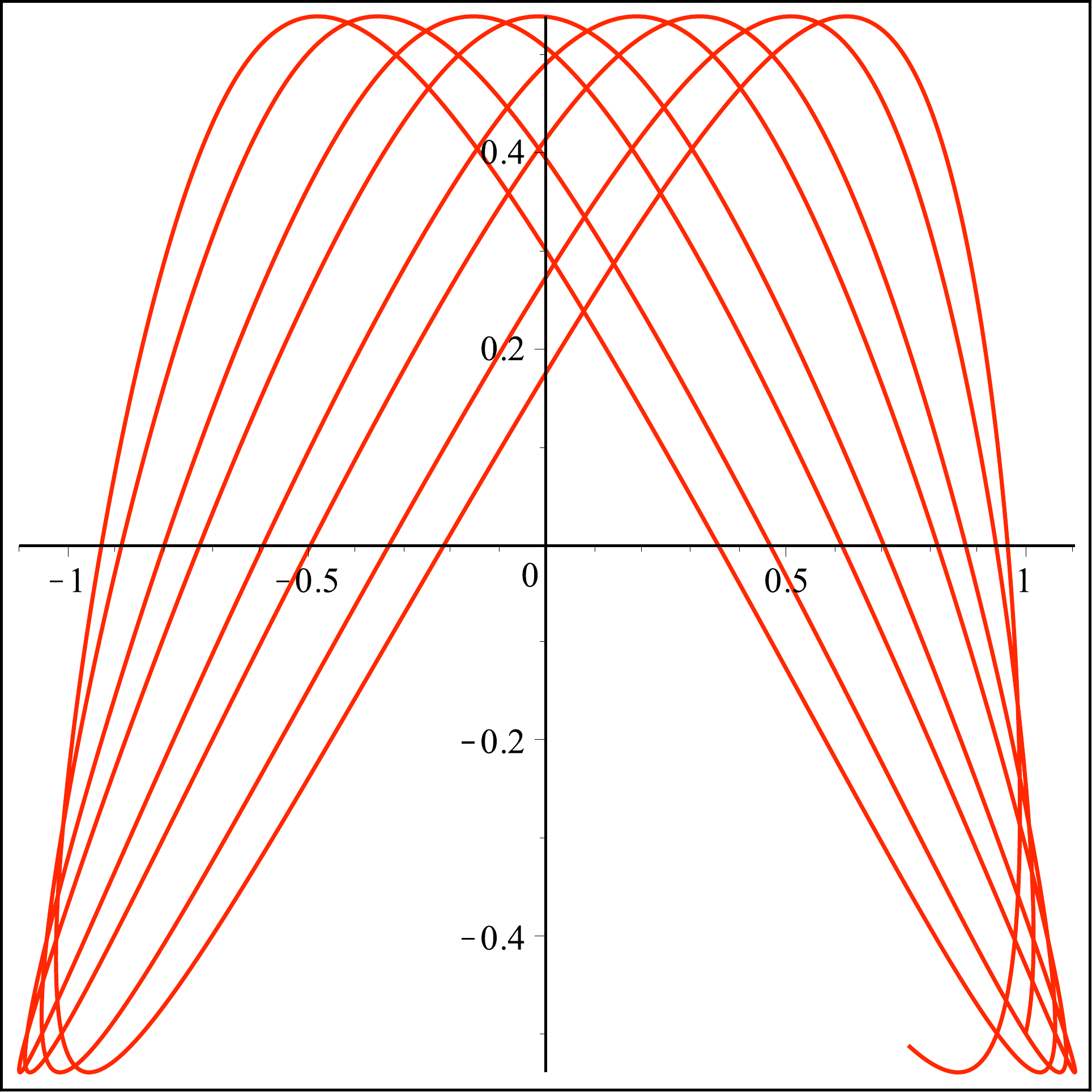}}
\put(79,72){(b)}
\put(11,0){\includegraphics[scale=0.3]{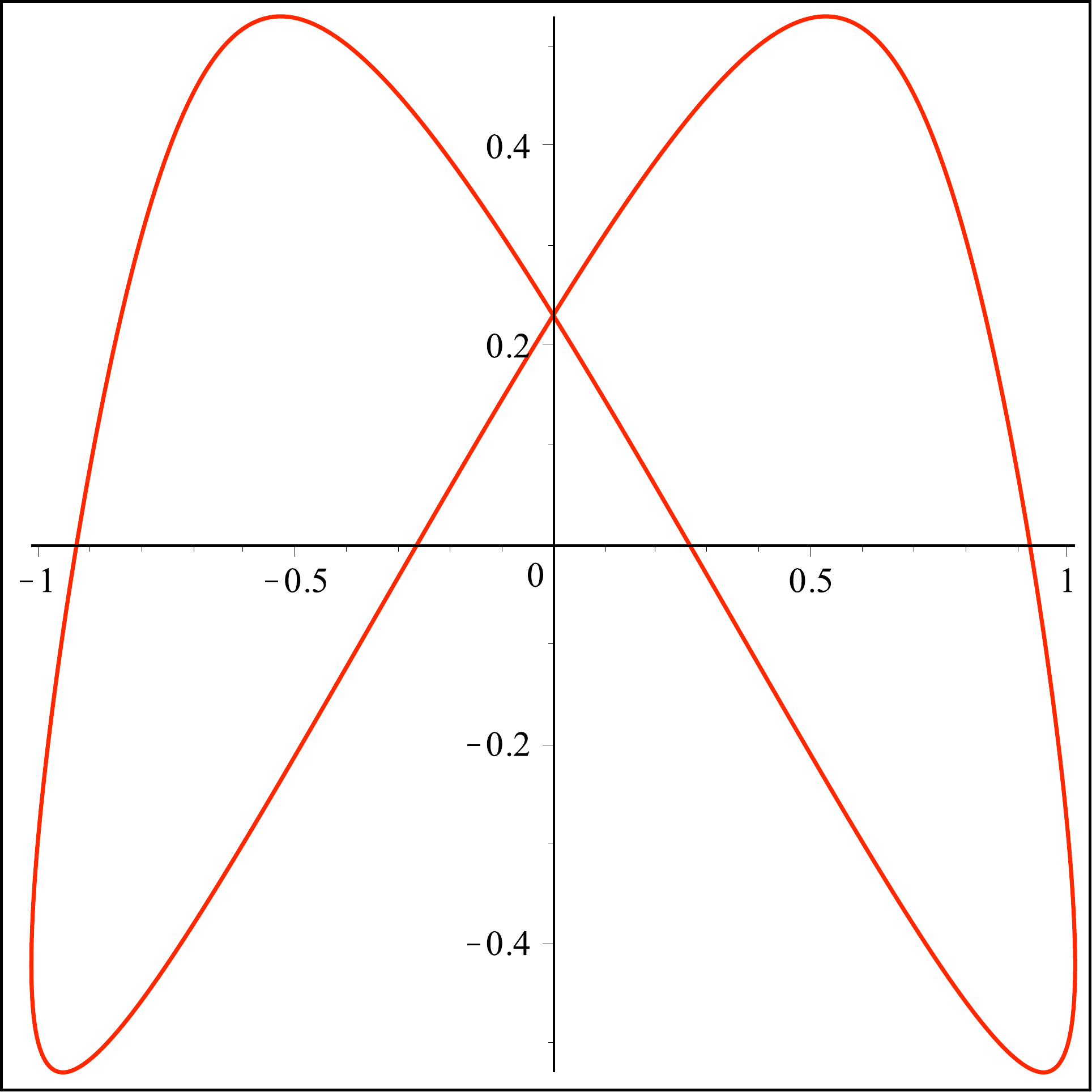}}
\put(5,3){(c)}
\put(85,0){\includegraphics[scale=0.3]{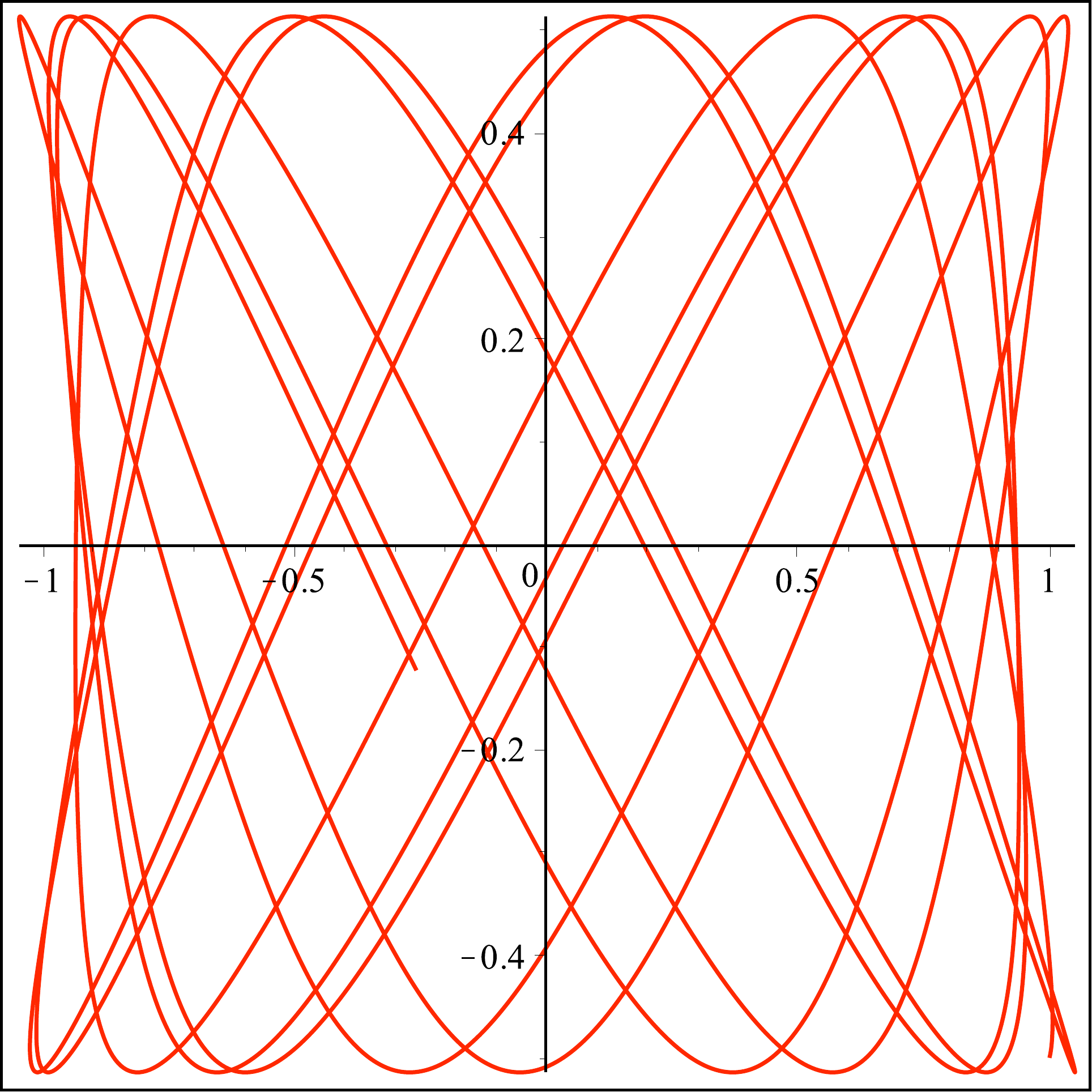}}
\put(79,3){(d)}
}
\end{picture}
\caption{ \footnotesize Trajectories on ${\mathbf S}^2$ corresponding to figure 3, but now plotted on the 2D projective plane in terms of Beltrami variables $(\tq_1,\tq_2)$:
  (a) $\Omega_2=1$ (the    $ 1:1$  case),
(b)     $\Omega_2=3$,
(c) $\Omega_2=4$  (the superintegrable  $1:2$  case),
 and 
(d)  $\Omega_2=9$  (the    $ 1:3$  case). The projective dynamics makes more apparent the fact that   the trajectories for the non-superintegrable cases are non-closed.}
\end{figure}
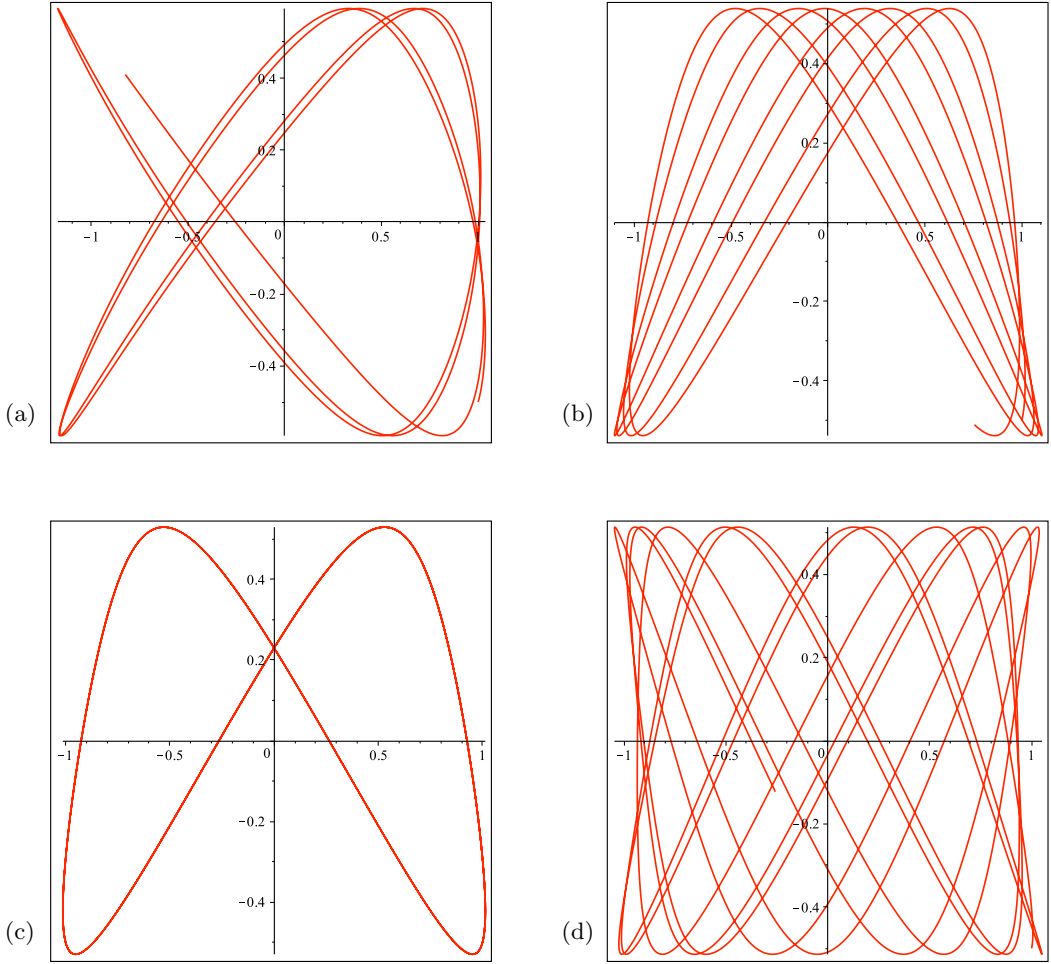


\begin{figure}[t]
\setlength{\unitlength}{1mm}
\begin{picture}(140,132)(0,0)
\label{figure6}
\footnotesize{
\put(11,69){\includegraphics[scale=0.3]{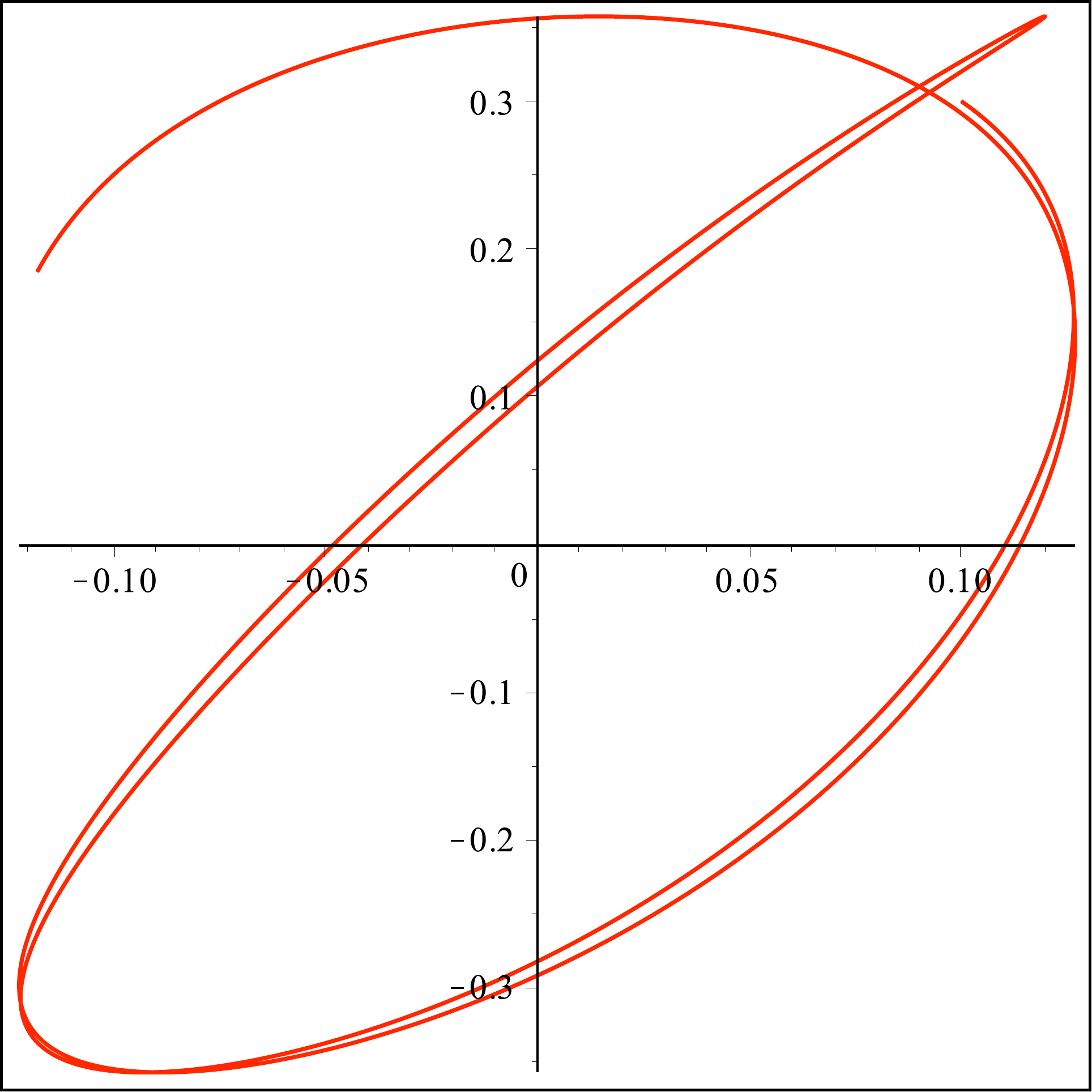}}
\put(5,72){(a)}
\put(85,69){\includegraphics[scale=0.3]{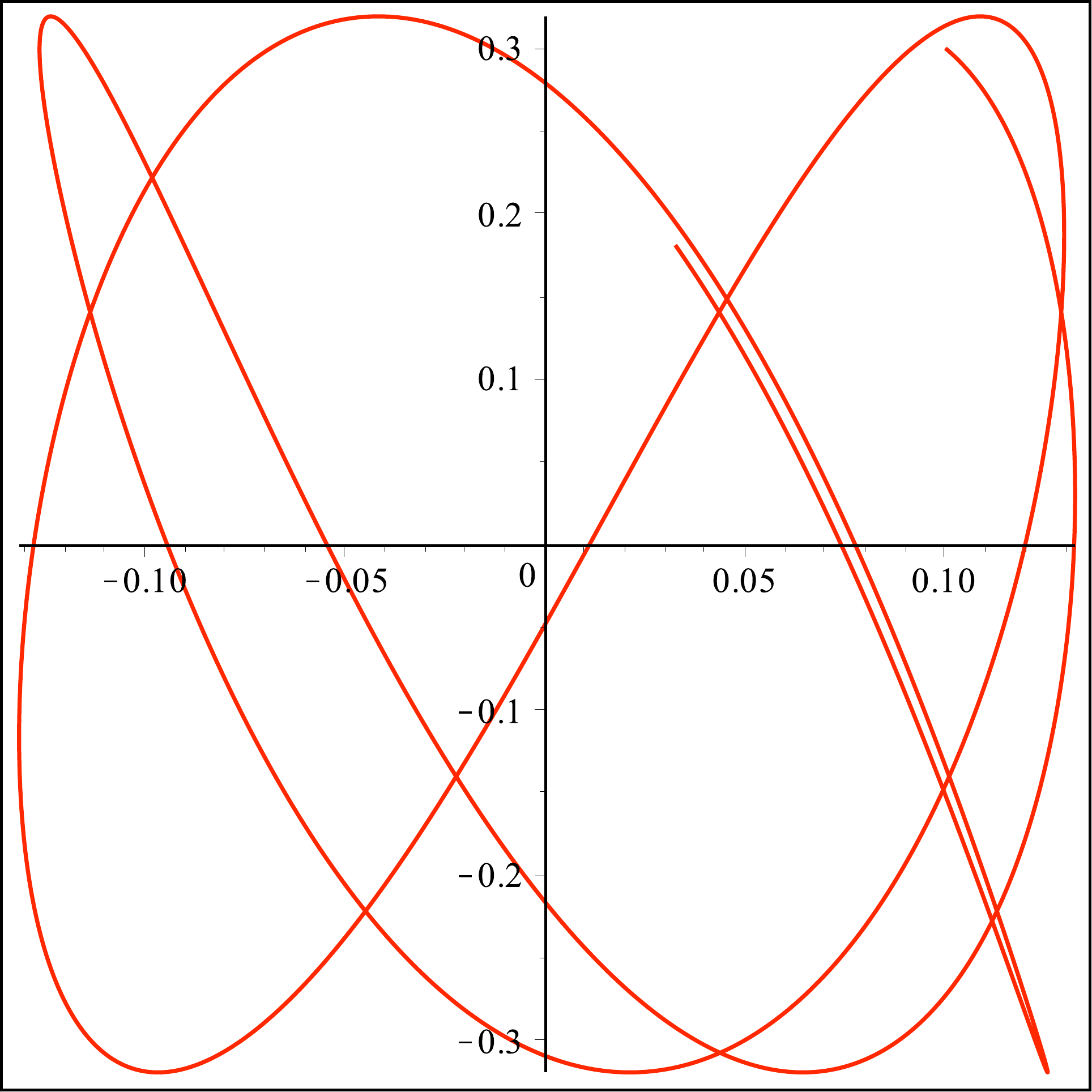}}
\put(79,72){(b)}
\put(11,0){\includegraphics[scale=0.3]{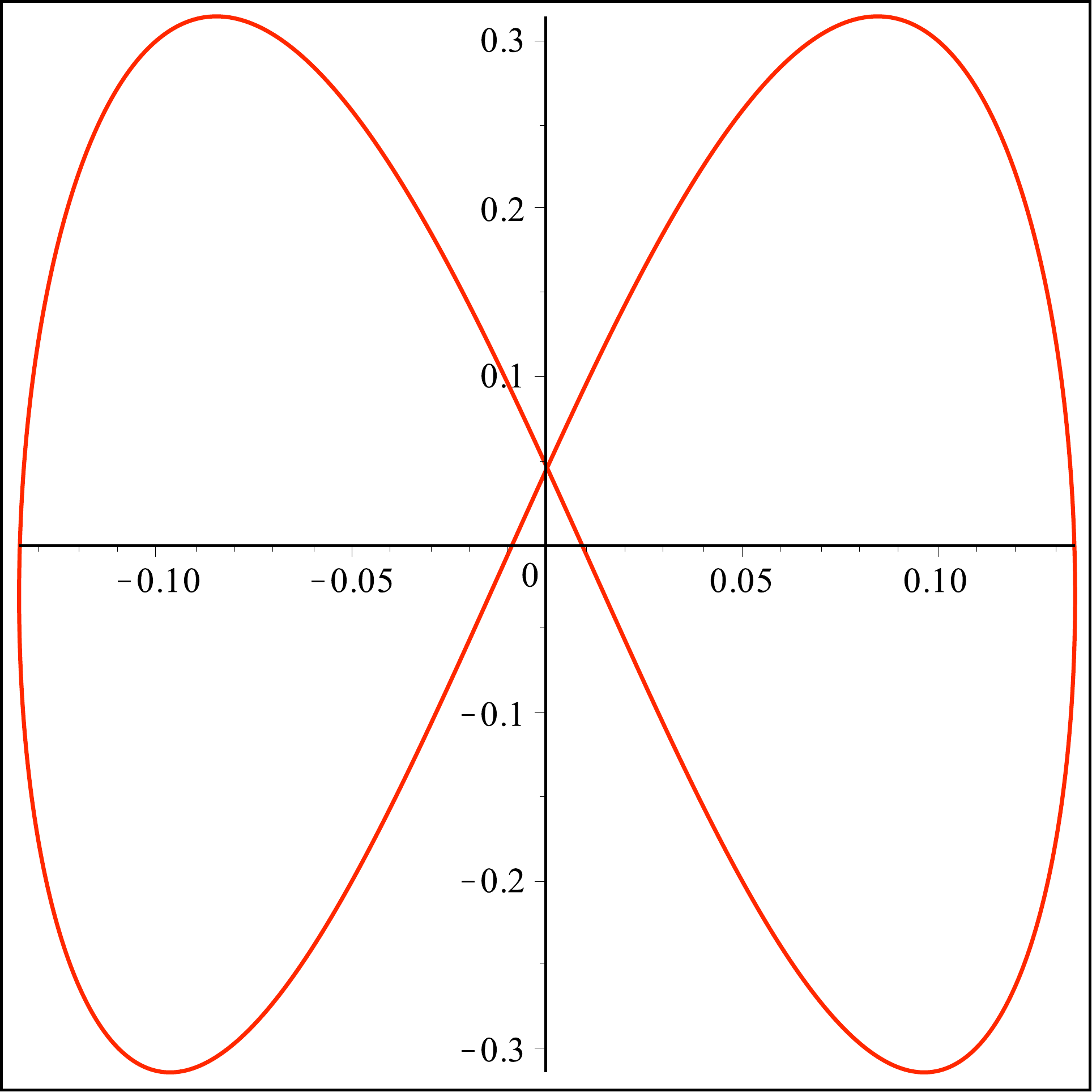}}
\put(5,3){(c)}
\put(85,0){\includegraphics[scale=0.3]{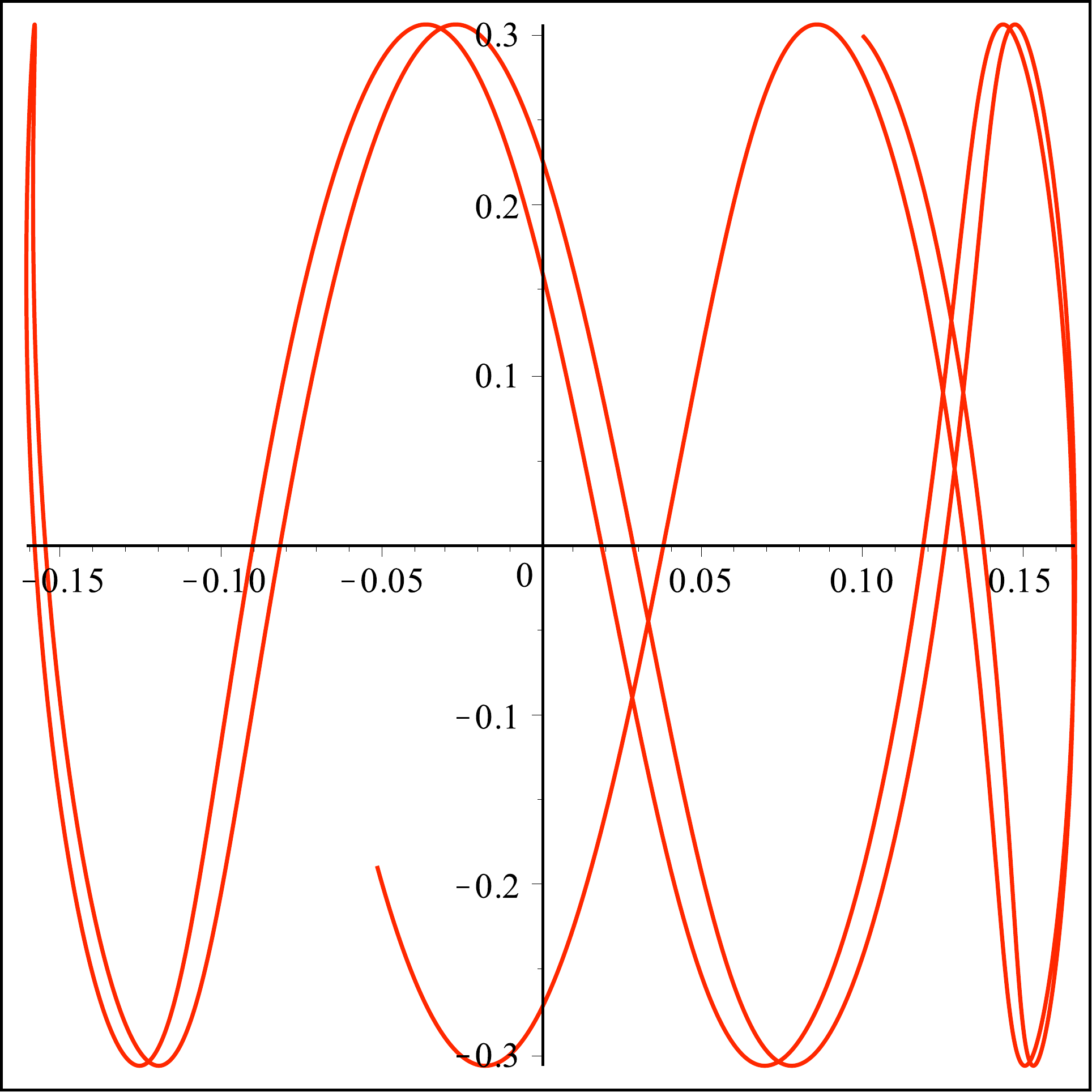}}
\put(79,3){(d)}
}
\end{picture}
\caption{ \footnotesize Trajectories on ${\mathbf H}^2$ from figure 4, but plotted on the 2D projective  plane  in terms of Beltrami variables $(\tq_1,\tq_2)$:
  (a) $\Omega_2=1$ (the     $1:1$  case),
(b)     $\Omega_2=3$,
(c) $\Omega_2=4$  (the superintegrable  $1:2$  case),
 and 
(d)  $\Omega_2=9$  (the    $ 1:3$  case). Again, the only closed trajectory is the one corresponding to the superintegrable (c) case.}
\end{figure}

We remark that   there is no clue in order to perform the explicit algebraic search of the possible additional integral for the $\nonn_\kk$  with $(\Om_1,\Om_2)$  associated to a commensurate Euclidean oscillator. Nevertheless, by taking into account that the superintegrability of a system implies that all its bounded trajectories are periodic (closed), we have performed a systematic numerical integration of $\nonn_\kk$ for many different values $(\Om_1,\Om_2)$ and several initial conditions for each of them. As a result of this numerical investigation, we find that the only closed trajectories for $\nonn_\kk$ are found for  $\Om_2=4\Om_1$ and $\la_2=0$, which corresponds to the quadratically superintegrable system described in Proposition 1.
Therefore, we conjecture that $\nonn_\kk$ is only integrable for generic values $(\Om_1,\Om_2)$.

 We present some of these trajectories for $\nonn_\kk$ on  $\>S^2$ and $\>H^2$ in figures 3 and 4, respectively. They are plotted  in ambient coordinates $(x_0,\>x)$, and  Rosochatius terms are neglected  ($\la_1=\la_2=0$). The numerical integration has been performed in Beltrami coordinates due to the simpler explicit form of (\ref{na}) with respect to (\ref{nc}). The initial Beltrami data have been chosen in such a manner that each trajectory is always confined in the hemisphere with $x_0>0$ in $\>S^2$, thus avoiding the problems with the equator $x_0=0$. Furthermore, the same trajectories are plotted in  figures 5 and 6, but now considering the projective plane with Beltrami coordinates $(\tq_1,\tq_2)$, with the same initial data as before.

As expected, a closed Lissajous-type $1:2$  curve only appears in figures 3(c),  4(c), 5(c) and 6(c) as a trajectory of the superintegrable $\non_\kk$ case with $(\Om_1=1,\Om_2=4)$. The remaining values of $(\Om_1,\Om_2)$ always provide non-periodic curves, despite they include the $(\Om_1=1,\Om_2=1)$ and $(\Om_1=1,\Om_2=9)$ cases, that do correspond to commensurate oscillators in the Euclidean limit. Note also that the projective Beltrami plots included in figures 5 and 6 allow one for a better numerical intuition about the non-superintegrability of the system.
    
Finally, we remark that the dynamical consequences of the Rosochatius $\la_i$-terms in~\eqref{na} can be numerically tested through a similar analysis to the one carried out in~\cite{Non} for the Hamiltonian~\eqref{Higgsall}. In particular, we find that the addition of the Rosochatius $\la_1$-term in~\eqref{na} does not alter the integrability properties of $\nonn_\kk$. When $\la_1>0$,  this term provides  a centrifugal barrier that restricts the configuration space, but the shape of the trajectories is the same and the only closed ones are those obtained in the superintegrable $(\Om_1=1,\Om_2=4)$ case. However, it is worth stressing that if we consider the second Rosochatius term ($\la_2\neq 0$) in~\eqref{na}, we find numerically that even the trajectories for the $(\Om_1=1,\Om_2=4)$ case are {\em non-closed}. This seems to indicate that the superintegrability of~\eqref{na} for $(\Om_1=1,\Om_2=4)$ is broken when the second Rosochatius term is added. 

We would like to stress that  this  is   a new  and somewhat unexpected result. In this respect, we recall  that if  the two Rosochatius terms are added to the 2D Higgs oscillator in the form~(\ref{Higgsall}),  the quadratic superintegrability of the system is preserved \cite{RS, int, CRMVulpi}.  In contrast, it is well known that the {\em quadratic} superintegrability of the Kepler--Coulomb system both in its (flat) Euclidean and curved versions on ${\mathbf S}^2$ and ${\mathbf H}^2$ is preserved  only if a single Rosochatius term is added~\cite{RS}. Nevertheless, when the second Rosochatius term is considered, the Euclidean and curved  Kepler--Coulomb Hamiltonians  have  been shown to be superintegrable, but  in this case with an `additional' {\em quartic} integral~\cite{Verrier,Kepler}. In this sense, one could expect that the full $(\Om_1=1,\Om_2=4)$ Hamiltonian~\eqref{na} with both Rosochatius potentials could be superintegrable with higher-order integrals, but this does not seem to be the case, and probably the anisotropy of the Hamiltonian plays a relevant role in this respect.


\sect{Concluding remarks and open problems}

The results here presented seem to support strongly the conjecture proposed in~\cite{Non}: for each commensurate $m:n$ Euclidean oscillator there should exist a different integrable anisotropic oscillator ${\cal H}_\k^{\Om_1,\Om_2}$ on ${\mathbf S}^2$ and ${\mathbf H}^2$ with arbitrary parameters $(\Om_1,\Om_2)$, and such that ${\cal H}_\k^{\Om_1,\Om_2}$ has the former Euclidean system as the zero curvature $\k\to 0$ limit when the $(\Om_1,\Om_2)$ parameteres are appropriately tuned to the $m:n$ commensurability condition. Moreover, when the $(\Om_1,\Om_2)$ parameters correspond to the $m:n$ condition, the Hamiltonian ${\cal H}_\k^{\Om_1,\Om_2}$ should be superintegrable, and its integrals of the motion should have the same degree in the momenta as the ones for the commensurate $m:n$ Euclidean oscillator.

So far, this conjecture has been proved to be valid for the $1:1$ system (the Higgs oscillator), whose anisotropic counterpart fulfilling the above conditions is~\eqref{Higgsall}, and for the $1:2$ system (the superintegrable oscillator~\eqref{MSall}), whose $(\Om_1,\Om_2)$ counterpart~\eqref{na} has been studied throughout  this paper. Obviously, in order to proceed with a rigorous proof of this statement, a generic expression for the curved superintegrable analogue of the Euclidean $m:n$ oscillator is needed, and this remains as a challenging open problem.

There are also two methodological aspects of the results we have obtained that we think deserve some attention. On one hand, the use of projective coordinates in this kind of algebraic integrability problems should be explored in more detail, since --as shown in figure 2-- it could provide a qualitative geometric insight into the specific features of superintegrable systems. On the other hand, the fact that by working appropriately with the curvature $\k$ as a `contraction' parameter, all the algebraic and geometric aspects of the transition between flat and curved dynamics become much more evident. In this framework, it is also worth mentioning that the Lorentzian geometry counterpart of the results here presented ({\em i.e.}, the corresponding integrable systems on the (1+1) dimensional (anti-)de Sitter and Minkowskian spacetimes) could be obtained from the results here presented by introducing appropriately a new contraction parameter that would be related with the speed of light~\cite{conf,kiev}. Work on all these lines is in progress.


\section*{Acknowledgments}

This work was partially supported by the Spanish MINECO under grants  MTM2010-18556 and AIC-D-2011-0711 (MINECO-INFN).


\end{document}